\documentclass[journal]{IEEEtran}

\ifCLASSINFOpdf

\else

\fi
\usepackage{cite}

\usepackage{color}

\usepackage{amsmath}
\usepackage{amssymb}
\usepackage{textcase}
\usepackage{soul}
\usepackage{indentfirst}
\usepackage{bm}
\usepackage{dsfont}
\usepackage[ruled,vlined]{algorithm2e}
\usepackage{siunitx}
\usepackage{multirow}
\usepackage{tabu}
\usepackage{multirow}
\usepackage{booktabs}
\usepackage{graphicx}
\usepackage[para,online,flushleft]{threeparttable}
\usepackage{lineno,hyperref}
\modulolinenumbers[5]

\ifCLASSOPTIONcompsoc
\usepackage[caption=false,font=normalsize,labelfon
t=sf,textfont=sf]{subfig}
\else
\usepackage[caption=false,font=footnotesize]{subfig}
\fi

\newcommand{\bmat}[1]{\begin{bmatrix} #1 \end{bmatrix}}

\newtheorem{theorem}{Theorem}
\newtheorem{definition}{Definition}
\newtheorem{remark}{Remark}

\begin{document}

\title{Vulnerability Analysis of Smart Grids\\ to GPS Spoofing}

\author{Paresh~Risbud,~\IEEEmembership{Student Member,~IEEE,}
        Nikolaos~Gatsis,~\IEEEmembership{Member,~IEEE,}
        and~Ahmad~Taha,~\IEEEmembership{Member,~IEEE}
\thanks{This material is based upon work supported by the National Science Foundation under Grant No. ECCS-1462404, and by a Smart Grid Security Research Grant from the UTSA OVPR. The authors are with the Dept.\ of Electrical \& Computer Eng., The Univ. of Texas at San Antonio. E-mails: \{Paresh.Risbud, Nikolaos.Gatsis, Ahmad.Taha\}@utsa.edu.}
}

\maketitle

\begin{abstract}
Sensors such as phasor measurement units (PMUs) endowed with GPS receivers are ubiquitously installed providing real-time grid visibility. A number of PMUs can cooperatively enable state estimation routines. However, GPS spoofing attacks can notably alter the PMU measurements, mislead the network operator, and drastically impact subsequent corrective control actions. Leveraging a novel measurement model that explicitly accounts for the GPS spoofing attacks, this paper formulates an optimization problem to identify the most vulnerable PMUs in the network. A greedy algorithm is developed to solve the aforementioned problem. Furthermore, the paper develops a computationally efficient alternating minimization algorithm for joint state estimation and attack reconstruction. Numerical tests on IEEE benchmark networks validate the developed methods.
\end{abstract}

\begin{IEEEkeywords}
GPS spoofing, PMU, State Estimation, Time Synchronization Attack, Weighted Least Squares.
\end{IEEEkeywords}

\maketitle

\IEEEpeerreviewmaketitle

\section*{Nomenclature}

\begin{description}[\IEEEsetlabelwidth{$V_{n,r},V_{n,i}$}\IEEEusemathlabelsep] 
{\item[$N_b$] Number of buses
\item[$N_l$] Number of transmission lines
\item[$L_n$] Number of lines connected to bus $n$
\item[$\mathcal{N}_n$] Set of buses connected to bus $n$
\item[$\mathcal{N}_{\text{PMU}}$] Set of buses with PMUs installations
\item[$N_p$] Number of PMUs attacked
\item[$\mathbf{a}$] Binary vector of PMU locations in a network
\item[$\mathbf{b}$] Binary vector of attacked PMUs
\item[$\mathbf{H}_n$] Regression matrix for bus $n$
\item[$\mathbf{w}_n$] Noise vector at bus $n$
\item[$\mathbf{\Sigma}_n$] Noise covariance for the measurement at bus $n$
\item[$V_n$] Voltage phasor at bus $n$
\item[$V_{n,r},V_{n,i}$] Real and imaginary parts of $V_n$
\item[$|V_n|,\theta_n$] Magnitude and angle of $V_n$
\item[$\mathbf{v}_r$] Vector collecting $V_{n,r}$ for all buses
\item[$\mathbf{v}_i$] Vector collecting $V_{n,i}$ for all buses
\item[$\mathbf{v}$] Vector  $[{\mathbf{v}_r}^{\top} \; {\mathbf{v}_i}^{\top}]^{\top}$
\item[$I_{nk}$] Current phasor on line $(n,k)$
\item[$|I_{nk}|, \theta_{I_{nk}}$] Magnitude and angle of $I_{nk}$
\item[$\hat{\mathbf{v}}_{\mathrm{ML}}$]  Maximum likelihood (ML) estimate of system state 
\item[$\mathbf{z}^{\mathrm{true}}_n$] Noiseless measurement at bus $n$
\item[$\mathbf{z}^{\mathrm{atk}}_n$] Noisy attacked measurement at bus $n$
\item[$\hat{\mathbf{v}}_{\mathrm{ML}}^{\mathrm{atk}}$] ML state estimate using attacked measurements
\item[$\bm\mu_{\mathrm{ML}}^{\mathrm{atk}}$] Expected value of $\hat{\mathbf{v}}_{\mathrm{ML}}^{\mathrm{atk}}$
\item[$\mathbf{B}_{\mathrm{ML}}$] Bias-scaling matrix induced by attack
\item[$\Delta\theta_n$] Attack angle at bus $n$
\item[$\bm{\Delta\theta}$] Vector collecting $\Delta\theta_n$ for all buses
\item[$\bm{\Gamma}_n$] Block diagonal matrix relating $\mathbf{z}^{\mathrm{true}}_n$ and $\mathbf{z}^{\mathrm{atk}}_n$
\item[$\bm\gamma_n$] Vector $\bmat{\cos(\Delta\theta_{n}) & \sin(\Delta\theta_{n})}^{\top}$
}
\end{description}

\section{Introduction}
\label{sec:intro}
\IEEEPARstart{P}{hasor} Measurement Units (PMUs) equipped with GPS receivers are installed ubiquitously in smart grids, replacing and augmenting traditional sensors of the Supervisory Control and Data Acquisition (SCADA) systems. The higher sampling rates of PMUs compared to SCADA systems assist the network operator to perform real time Wide Area Monitoring, Protection and Control (WAMPAC)---a unique feature of smarter power grids.

Cyber attacks on PMUs include False Data Injection (FDI) attacks and GPS spoofing attacks. In an FDI attack, the attacker tries to inject false data that are not detectable by bad data detection algorithms into the network; see e.g., the recent works \cite{J.Liang,Gao2016}. GPS spoofing is caused by transmitters mimicking the GPS signal with the intention of altering the GPS time estimated by the PMU's GPS receiver~\cite{Kumar,Jafarnia-Jahromi}. These attacks maliciously introduce erroneous time stamps, thereby inducing a wrong phase angle in the PMU measurements \cite{Schmidt}, and are also called time synchronization attacks (TSAs).

The Electric Power Research Institute, in collaboration with the National Electric Sector Cybersecurity Organization Resource, has published a technical report recognizing the vulnerability of PMUs to GPS spoofing under its scenario \textit{WAMPAC.12: GPS Time Signal Compromise} \cite{NESCOR}.   
This paper considers the effects of GPS spoofing attacks on power grids.

Sensors such as PMUs cannot be placed at every bus location in the power network as they are still costly. Thus they are placed in crucial network locations to ensure the visibility of the network. Consequently, a system operator can generate information about other, \textit{non-sensed} states using state estimation (SE) routines \cite{Ali_Abur,GGVKNG-spmag,Taha2015,Hu}. The purpose of SE is three-fold: protection, operational control, and grid status verification \cite{Taha2015}. In a network, measurements from dispersed PMUs are time synchronized at the Phasor Data Concentrator (PDC) and fed to the super PDC for further processing. Thus, systematic attacks on PMUs can have a catastrophic effect on the SE and consequently on the grid.

This paper analyzes the vulnerability of smart grids to GPS spoofing by formulating an optimization problem to identify vulnerable PMUs in the network, and by developing an algorithm for joint state estimation and attack reconstruction. { The ensuing section provides a review of the related literature, and details the contributions of this paper.}

\section{Literature Review \& Paper Contributions}
\label{sec:Literature}

This section categorizes the GPS spoofing attack literature in three groups, namely, the feasibility of spoofing attacks, impacts to the power grid, and countermeasures, and details the contributions of this paper. 

The \textbf{experimental and theoretical feasibility} of GPS spoofing attacks is shown in 
\cite{Humphreys2008, Bonebrake, Shepard2012, Akkaya,  Jiang}.
More specifically, \cite{Humphreys2008} demonstrates the implementation of a laboratory-based GPS spoofer and discusses mechanisms against civilian GPS spoofing. A spoofer implemented in US DOE's Pacific Northwest National Laboratory is the theme of \cite{Bonebrake}. The vulnerabilities of PMUs to GPS spoofing are studied in \cite{Shepard2012} and \cite{Akkaya}. 
The work in \cite{Jiang} manipulates the GPS navigation data to achieve various timing and phase errors in PMU measurements. 

The \textbf{impacts} of GPS spoofing attacks on system operations are studied in \cite{Zhang, Jiang, Shepard2012}. 
Specifically, the effect of GPS spoofing on fault location and voltage stability monitoring algorithm is demonstrated in \cite{Jiang} and \cite{Zhang}. A falsely activated generator trip scenario---as a result of a GPS spoofing attack---is constructed in~\cite{Shepard2012}. In our previous work \cite{Risbud2016}, the statistics of state estimates affected by GPS spoofing are studied.

\textbf{Countermeasures} to spoofing attacks are discussed in 
\cite{Fan2015, Arvani2014, Yu2014, Psiaki2016, Zhu2016, Mousavian2015}.
A detection method for TSAs on multiple PMUs using a cross layer detection mechanism is developed in \cite{Fan2015}.
 The work in \cite{Arvani2014} presents a method to detect hazardous data via signal-based and model-based methods. Furthermore, a GPS spoofing attack detection scheme based on collaboration among multiple PMUs in a large grid is the theme of \cite{Yu2014}. Recommendations for improving spoofing detection in commercial receivers are articulated in \cite{Psiaki2016}. The work in \cite{Zhu2016} takes advantage of the PMU locations along with the statistics of GPS receivers to detect the spoofing attacks. The work in \cite{Mousavian2015} proposes methods to prevent propagation of cyber-attacks in PMU networks.

 State estimation and GPS spoofing attack identification in power networks is pursued in \cite{Fan2017,Pradhan2016ConfCNS}. A computationally efficient algorithm to identify an attack on at most one PMU is developed in~\cite{Fan2017}. The generalized likelihood ratio test (GLRT) is the basis of the method developed in~\cite{Pradhan2016ConfCNS}, but implementation of GLRT requires the solution of a hard optimization problem with respect to the unknown time delay induced by GPS spoofing. Beyond the two aforementioned works, which specialize in GPS spoofing, recent works that deal with attack identification in power networks include~\cite{Gao2016, Taha-RiskMitigation, forti2016bayesian, hu2017secure}. The majority of these studies require successive measurements across time and detailed dynamical models of the network, including generator parameters. The contributions of this paper are as follows.
\begin{itemize}
\item A measurement model that explicitly relates the PMU measurements with the network state and the (potential) GPS spoofing-induced phase shifts is developed. This model is leveraged to derive the statistics of the state estimator when GPS-spoofed PMU measurements are utilized.
\item A novel greedy algorithm to identify the most vulnerable PMU locations is developed. The vulnerability is quantified by the state estimation error, which is characterized explicitly in terms of the spoofing attack. Exhaustive search is used to validate the performance of the greedy algorithm. Our analysis is intended to guide the system operator towards increased protection of the network's most vulnerable PMU locations. 
\item An algorithm to jointly perform state estimation and attack angle reconstruction is developed. The algorithm is based on alternating minimization of a bilinear least squares objective. Lagrangian duality is leveraged so that---despite the nonconvexity---the algorithm features closed-form updates. This property renders the algorithm attractive for real-time implementation.  
\item The algorithm is extended to state estimation and attack identification using combined SCADA and GPS-spoofed PMU measurements.
\end{itemize}

Numerical tests indicate that the alternating minimization algorithm can yield smaller state estimation error than the largest normalized residual test (LNRT), which is a classical method  for bad data identification~\cite[Sec.~4.8.4]{Antonio2008}. The algorithm is also compared to one of the previous GPS spoofing identification approaches~\cite{Fan2017}. The developed algorithm can identify simultaneous attacks to more than one PMU, in contrast to~\cite{Fan2017}. Even if 20\% of the PMUs are attacked, the algorithm correctly identifies the attacked PMU locations and yields an accurate state estimate, as indicated by tests on standard IEEE transmission networks.

In comparison with~\cite{Pradhan2016ConfCNS},  this work does not necessitate the solution of a complicated optimization problem with respect to the induced phase shift. The state estimation algorithm is based on a new measurement model that explicitly relates the PMU measurements with the attacked phases, which is different than the additive attack model of~\cite{Gao2016, Taha-RiskMitigation, forti2016bayesian, hu2017secure}. The implication is that the developed algorithm does not require successive measurements across time, and is also able to identify the spoofing-induced attacked phases.

The remainder of the paper is organized as follows. Section~\ref{sec:se} reviews the PMU-based SE. Section~\ref{sec:Attack_Formulation} formulates an optimization problem to identify the most vulnerable PMUs in the network. An algorithm for joint state estimation and attack reconstruction is developed in Section \ref{sec:AM}. Extensions to SE with PMU and SCADA measurements are discussed in~\ref{sec:PMU_SCADA}. Numerical tests on standard IEEE networks are performed in Section \ref{sec:Num_tests}, and Section \ref{sec:Conclusion} concludes the paper. 

\section{PMU-Based State Estimation}
\label{sec:se}
This section outlines the network and measurement model, with and without TSAs, and provides the optimal state estimators using PMU measurements without TSAs.
\subsection{Network Model and PMU Measurements}
\label{subsec:net}
Consider a power network with $N_b$ buses connected via $N_l$ transmission lines. Let $\mathcal{N}_n$ be the set of buses connected to bus~$n$, and define $L_n=|\mathcal{N}_n|$ as the number of lines connected to bus~$n$. The  system state is the vector of nodal voltages in rectangular coordinates denoted by $\mathbf{v} = [{\mathbf{v}_r}^{\top} \; {\mathbf{v}_i}^{\top}]^{\top}\in \mathds{R}^{2N_b \times 1}$ where $\mathbf{v}_{r}$ and $\mathbf{v}_{i}$ collect the real and imaginary parts $V_{n,r}$ and  $V_{n,i}$  of the complex voltages at buses $n=1,\ldots,N_b$.

PMUs are installed on select buses of the network; $a_n$ is a binary indicator is equal to 1 if a PMU is installed at bus $n$ and 0 otherwise. Vector $\mathbf{a}$ collects $a_n$ for $ n = 1, 2, \dotsc , N_b$. The set of buses where PMUs are  installed is denoted by 
$\mathcal{N}_{\mathrm{PMU}} = \{ i \in \{ 1,2,\ldots,N_b\}| {a_i = 1}  \}.$
A PMU installed at bus $n$ measures the bus's complex voltage as well as the complex currents on all lines that bus $n$ is connected to. This collection of measured quantities (in rectangular coordinates) at bus $n$ is concatenated in a vector $\mathbf{z}_n\in\mathds{R}^{2+2L_n}$. To make the notation more compact, define $M_n=2+2L_n$ as the number of distinct real quantities measured by the PMU at bus $n$.

Let $V_n$ and $I_{nk}$ generically denote the voltage and current phasors at bus $n$ and line $(n,k)$ respectively; and let $\theta_n$ and $\theta_{I_{nk}}$ denote the corresponding phasor angles. 
It is convenient for subsequent developments to consider the noiseless version of $\mathbf{z}_n$, which is denoted by $\mathbf{z}_n^{\mathrm{true}}\in\mathds{R}^{M_n}$:
\begin{equation} \label{eq:zntruedef}
\mathbf{z}^{\mathrm{true}}_n = \bmat{V_{n,r} \\ V_{n,i} \\ 
\{I_{nk,r}\}_{k\in\mathcal{N}_n}\\ \{I_{nk,i}\}_{k\in\mathcal{N}_n}} = \bmat{|V_{n}| \cos(\theta_n) \\ |V_{n}| \sin(\theta_n) \\ 
\{|I_{nk}|\cos(\theta_{I_{nk}})\}_{k\in\mathcal{N}_n}\\ \{|I_{nk}|\sin(\theta_{I_{nk}})\}_{k\in\mathcal{N}_n} }
\end{equation}
where $I_{nk,r}$ and  $I_{nk,i}$ are the real and imaginary parts of the complex current injected into line $(n,k)$. 
Note that the current injected into line $(n,k)$ is different than the current injected into line $(k,n)$. 
To summarize, the noiseless quantities measured at bus $n\in\mathcal{N}_{\mathrm{PMU}}$ comprise the real and imaginary parts of the nodal complex voltage, appended by the real and imaginary parts of the complex currents injected to all lines connected to bus $n$. Using the bus admittance matrix of the network, $\mathbf{z}_n^{\mathrm{true}}$ can be written as a linear function of the system state $\mathbf{v}$ as
$ \mathbf{z}_{n}^{\mathrm{true}} = \mathbf{H}_{n} \mathbf{v}$. 
The construction of $\mathbf{H}_{n} \in \mathds{R}^{M_n\times 2N_b}$ is provided in \cite{Kekatos,Risbud2016}. In practice, a PMU at bus $n$ measures $\mathbf{z}_n$, which is a noisy version of $\mathbf{z}_n^{\mathrm{true}}$, i.e., $$\mathbf{z}_{n}=  \mathbf{z}_{n}^{\mathrm{true}} + \mathbf{w}_n=  \mathbf{H}_{n} \mathbf{v} + \mathbf{w}_n$$ where $\mathbf{w}_n \sim \mathcal{N}(0,\bm{\Sigma}_n)$ represents an additive Gaussian noise vector that is assumed independent across PMUs and has a known positive definite covariance $\bm\Sigma_n$. 
	  
Given that the likelihood of the measurement $p\bigl(\{\mathbf{z}_{n}\}_{n=1}^{N_b};\mathbf{v}\bigr)$  = 
$ \prod_{n=1}^{N_b} p(\mathbf{z}_{n};\mathbf{v})^{a_n}$ is Gaussian, the maximum likelihood (ML) estimate of the system state is given as
\begin{equation} \label{eq:mlecost}
\hat{\mathbf{v}}_{\mathrm{ML}} = \underset{\mathbf{v}}{\mathrm{arg min}} \sum_{n=1}^{N_b}a_n(\mathbf{z}_n - \mathbf{H}_{n}\mathbf{v})^{\top}\bm{\Sigma}_n^{-1}(\mathbf{z}_n - \mathbf{H}_{n}\mathbf{v}).
\end{equation}
The optimization in \eqref{eq:mlecost} amounts to unconstrained least squares, and can be solved by taking the gradient of the cost function with respect to $\mathbf{v}$ and setting it to zero, resulting in
\begin{equation} \label{eq:mle} 
\hat{\mathbf{v}}_{\mathrm{ML}} = \mathbf{G}^{-1}\sum_{n=1}^{N_b}a_n\mathbf{H}_n^{\top}\bm{\Sigma}_n^{-1}\mathbf{z}_n
\end{equation}
where it is assumed that the matrix $\mathbf{G}= \sum_{n=1}^{N_b}a_n\mathbf{H}_n^{\top}\bm{\Sigma}_n^{-1}\mathbf{H}_n$
is non-singular. Invertibility of $\mathbf{G}$ is tantamount to the state observability, which can be ensured when there is a sufficient number of installed PMUs in the network~\cite{Kekatos}.

Substituting $\mathbf{z}_{n}=\mathbf{H}_{n} \mathbf{v} + \mathbf{w}_n$ into~\eqref{eq:mle} yields the statistics of the estimator as $\hat{\mathbf{v}}_{\mathrm{ML}} \sim \mathcal{N}(\mathbf{v},\mathbf{G}^{-1})$. That is, the expected value of the estimate is the system state $\mathbf{v}$, or in other words the estimator is unbiased. The next section develops a relationship between the measured quantities of bus $n$ when there is TSA and their version in the absence of an attack.

\subsection{TSA-Impacted PMU Measurement Model}\label{subsec:TSAmodel}
 As mentioned in \cite{Zhang}, TSA affects only the phase of the measurement. Specifically, a TSA on bus $n$ introduces a clock offset error $\Delta t_n$ \cite{Zhang}. The phase angle error corresponding to the clock offset error for the PMU at bus $n$ is denoted by $\Delta \theta_n$ and is given by 
$\Delta\theta_n = 2 \pi f \Delta t_n [\text{rad}] = 360 f \Delta t_n [\text{degrees}]$
where $f = 60 $ Hz.
Table \ref{table:Phase} lists typical phase angle errors caused by GPS spoofing attacks reported in the literature.
\begin{table}[!t]
\caption{GPS-Spoofing induced phase angle error}
\label{table:Phase}
 \centering
  \begin{threeparttable}
	  \begin{tabular}{c|c}
	    \hline
		 Reference  & Phase angle error \\ 
		\hline
		\cite{Jiang}  & $52^{\circ}$ \\ 
		\cite{Shepard2012} & $70^{\circ}$\\ 
		\cite{Zhang} & $\pm 60^{\circ}$  \\
		\hline
	  \end{tabular} 
  \end{threeparttable}
\end{table}

The noisy attacked PMU measurement at bus $n$ is given by
\begin{equation} \label{eq:znatk}
 \mathbf{z}^{\mathrm{atk}}_n = \bmat{|V_n| \cos(\theta_n + \Delta{\theta_n}) \\ |V_n| \sin(\theta_n + \Delta{\theta_n}) \\   
 \{|I_{nk}| \cos(\theta_{I_{nk}} + \Delta{\theta_n})\}_{k\in\mathcal{N}_n}  \\
 \{|I_{nk}| \sin(\theta_{I_{nk}} + \Delta{\theta_n})\}_{k\in\mathcal{N}_n}
 } + \mathbf{w}_n
\end{equation} 
Note that there is a potentially different $\Delta\theta_n$ per bus $n$. Combining \eqref{eq:znatk} with~\eqref{eq:zntruedef} and introducing  $\mathbf{z}_{n}^{\mathrm{true}} = \mathbf{H}_{n} \mathbf{v}$, a linear relationship between $\mathbf{z}^{\mathrm{atk}}_n$ and $\mathbf{z}_n^{\mathrm{true}}$ can be derived as: 
\begin{align}\label{eq:znmeas}
	\mathbf{z}^{\mathrm{atk}}_n = \bm{\Gamma}_n \mathbf{z}_n^{\mathrm{true}} + \mathbf{w}_n = \bm{\Gamma}_n \mathbf{H}_n \mathbf{v} + \mathbf{w}_n
\end{align}  
where $\bm{\Gamma}_n\in\mathds{R}^{M_n \times M_n}$ is a block diagonal matrix\footnote{Note that $\bm\Gamma_n$ depends on $\Delta\theta_n$, but this dependency is kept implicit in order to keep the notations compact. When $\Delta\theta_n=0$ (no attack), we obtain the identity matrix, $\bm\Gamma_n=\mathbf{I}_{M_n}$.} consisting of $1+L_n$ blocks and each block is the $2\times 2$ matrix $\left[ \begin{smallmatrix} \cos\Delta{\theta_n} & -\sin\Delta{\theta_n}  \\ \sin\Delta{\theta_n} & \cos\Delta{\theta_n} \end{smallmatrix} \right]$.
Therefore, $\mathbf{z}_n^{\mathrm{true}}$ can be thought of as the noiseless measurement that would be available to the PMU-enabled bus $n$ in the absence of a spoofing attack. 
The measurement model in \eqref{eq:znmeas} connects the state with the attack. It is worth noticing that the measurement is  linear in the state, but nonlinear in the attack angle. 
In Section \ref{sec:Attack_Formulation}, we use the measurement model \eqref{eq:znmeas} to develop a framework which identifies the most vulnerable PMU locations in the network. In Section \ref{sec:AM}, the model in \eqref{eq:znmeas} is leveraged to jointly perform the state estimation and attack angle reconstruction from spoofed PMU measurements.

\section{Identifying Susceptible PMU Locations}\label{sec:Attack_Formulation}
This section formulates an optimization problem to identify the most vulnerable PMUs in the network and develops a computationally attractive solution algorithm. To this end, the statistics of the ML state estimate obtained from the corrupted measurement~\eqref{eq:znmeas} are characterized next.

\subsection{Statistics of the Estimates Under Attack}\label{subsec:statsSE} 
The worst-case scenario  is considered, whereby the SE routine does not know (or has not detected) that an attack has occurred. Under this scenario, the SE routine passes the corrupted measurement~\eqref{eq:znmeas} through the estimator~\eqref{eq:mle}. Thus, the state estimate after the attack is given by 
\begin{equation}
\label{eq:mlatk}
\hat{\mathbf{v}}_{\mathrm{ML}}^{\mathrm{atk}} = \textbf{G}^{-1}\sum_{n=1}^{N_b}a_n\mathbf{H}_n^{\top}\bm{\Sigma}_n^{-1}\mathbf{z}_n^{\mathrm{atk}}. 
\end{equation}  
It should be emphasized that~\eqref{eq:mlatk} is \emph{not} the ML estimate that the SE module would derive if the attack magnitude $\Delta\theta_n$ were known. With this observation, we refer to~\eqref{eq:mlatk} as the attacked ML estimator. In what follows, we derive the statistics of the estimates under attack.

\begin{theorem}\label{prop:MLstats}
The attacked ML estimator has the Gaussian distribution
$\hat{\mathbf{v}}_{\mathrm{ML}}^{\mathrm{atk}} \sim \mathcal{N}(\bm\mu_{\mathrm{ML}}^{\mathrm{atk}},\mathbf{G}^{-1})$, i.e., it has covariance $\mathbf{G}^{-1}$ and expected value
\begin{equation}\label{eq:meanMLatk}
\bm\mu_{\mathrm{ML}}^{\mathrm{atk}} = \mathbf{G}^{-1}\sum_{n=1}^{N_b}a_n\mathbf{H}_n^{\top}\bm{\Sigma}_n^{-1} \bm{\Gamma}_n \mathbf{H}_n \mathbf{v}.
\end{equation}
\end{theorem}

\begin{IEEEproof}
Substituting~\eqref{eq:znmeas} into~\eqref{eq:mlatk} yields
\begin{align*}
		E[\hat{\mathbf{v}}_{\mathrm{ML}}^{\mathrm{atk}}] &= 
	        E\left[\mathbf{G}^{-1}\left(\sum_{n=1}^{N_b}a_n\mathbf{H}_n^{\top}\bm{\Sigma}_n^{-1}(\bm{\Gamma}_n \mathbf{H}_n \mathbf{v} + \mathbf{w}_n)\right)\right]\\
		&=E\left[\mathbf{G}^{-1}\sum_{n=1}^{N_b}a_n\mathbf{H}_n^{\top}\bm{\Sigma}_n^{-1} \bm{\Gamma}_n \mathbf{H}_n \mathbf{v}\right]
\end{align*}
which proves~\eqref{eq:meanMLatk}.
Using the definition of covariance and the fact that $E[\mathbf{w}_n \mathbf{w}_n^{\top}]=\bm\Sigma_n$, it follows that
\begin{align*}
		\text{cov}[\hat{\mathbf{v}}_{ML}^{\mathrm{atk}}] &= E[(\hat{\mathbf{v}}_{\mathrm{ML}}^{\mathrm{atk}}-E[\hat{\mathbf{v}}_{\mathrm{ML}}^{\mathrm{atk}}])(\hat{\mathbf{v}}_{\mathrm{ML}}^{\mathrm{atk}}-			    E[\hat{\mathbf{v}}_{\mathrm{ML}}^{\mathrm{atk}}])^{\top}]\\
		&\mspace{-50mu}= \mathbf{G}^{-1}\left(\sum_{n=1}^{N_b}a_n^{2}\mathbf{H}_n^{\top}\bm{\Sigma}_n^{-1}E[\mathbf{w}_n \mathbf{w}_n^{\top}]\bm{\Sigma}_n^{-1} \mathbf{H}_n\right) ({\mathbf{G}^{-1}})^{\top}
\end{align*}
which verifies that $\text{cov}[\hat{\mathbf{v}}_{\mathrm{ML}}^{\mathrm{atk}}]=\mathbf{G}^{-1}$.
\end{IEEEproof}
Having derived the distribution of the attacked estimate, the next section details how estimation accuracy metrics typically employed in power systems are affected by the attack.

\subsection{SE Accuracy Metrics}
For any estimator $\hat{\mathbf{v}}$, the mean square error (MSE) matrix is a metric of the estimation accuracy which is introduced in the power systems as early as the seminal work in~\cite{Schweppe-PartI} and has traditionally been utilized in the statistical literature~\cite{kay1993fundamentals}.
\begin{definition}
\label{def:msem}
With $\mathbf{v}$ denoting the true system state, the mean square error (MSE) matrix is defined by
\begin{equation}
\mathsf{MSEM}(\hat{\mathbf{v}}) = E\left[(\hat{\mathbf{v}} - \mathbf{v})(\hat{\mathbf{v}} - \mathbf{v})^\top\right].
\label{eq:MSEM}
\end{equation}
\end{definition}
The MSE matrix is formed by the pairwise nodal voltage estimate errors.  Its diagonal entries characterize the accuracy of individual nodal voltage estimates. Based on the MSE matrix,  it is customary to obtain a scalar metric of the SE accuracy as follows.
\begin{definition}
\label{def:mse}
The mean square error of an estimator is defined as the trace of the MSE matrix:
\begin{equation}
\mathsf{MSE}(\hat{\mathbf{v}}) = \mathsf{trace}[\mathsf{MSEM}(\hat{\mathbf{v}})]= E\left[(\hat{\mathbf{v}} - \mathbf{v})^\top(\hat{\mathbf{v}} - \mathbf{v})\right].
\label{eq:MSE}
\end{equation}
\end{definition}
The MSE sums the squared errors of nodal voltage estimates. The last equality in~\eqref{eq:MSE} follows immediately from the linearity of the expectation. The MSE matrix and the MSE are formulated by squaring the difference between the estimate and its true value. This difference yields a bias metric as follows.
\begin{definition}
\label{def:bias}
The bias of an estimator is defined as
\begin{equation}
\mathsf{Bias}(\hat{\mathbf{v}})  = E[\hat{\mathbf{v}} - {\mathbf{v}}] = E[\hat{\mathbf{v}}] - {\mathbf{v}}.
\label{eq:bias}
\end{equation}
\end{definition}

Theorem~\ref{prop:MLstats} reveals that GPS spoofing introduces a bias in the ML estimator, and is given by 
$\mathsf{Bias}(\hat{\mathbf{v}}_{\mathrm{ML}}^{\mathrm{atk}}) = \bm\mu_{\mathrm{ML}}^{\mathrm{atk}} - \mathbf{v} = \mathbf{B}_{\mathrm{ML}}(\bm{\Delta\theta}) \mathbf{v}$
where $\bm{\Delta\theta}= [\Delta\theta_1, \ldots, \Delta\theta_{N_b}]^T$ and 
\begin{equation}
\mathbf{B}_{\mathrm{ML}}(\bm{\Delta\theta})  = \mathbf{G}^{-1}\sum\nolimits_{n=1}^{N_b}a_n\mathbf{H}_n^{\top}\bm{\Sigma}_n^{-1} \bm{\Gamma}_n(\bm{\Delta\theta}) \mathbf{H}_n - \mathbf{I}.
\end{equation}
The matrix $\mathbf{B}_{\mathrm{ML}}(\bm{\Delta\theta})$ analytically characterizes the effect of the attacked PMU  phase shift to the estimation bias. The ideal scenario for the bias is to be zero, which occurs in the absence of an attack, yielding $\bm{\Delta\theta}=\mathbf{0}$ and $\mathbf{B}_{\mathrm{ML}}(\bm{\Delta\theta})=\mathbf{0}$. 

The next theorem precisely characterizes the relationship between the MSE and the bias.
\begin{theorem}
\label{thm:MSE}
The MSE and the bias of the attacked state estimate $\hat{\mathbf{v}}_{\mathrm{ML}}^{\mathrm{atk}}$ satisfy the following relationship:
\begin{equation}
\mathsf{MSE}\left(\hat{\mathbf{v}}_{\mathrm{ML}}^{\mathrm{atk}}\right) = \mathsf{trace}\left[\mathbf{G}^{-1}\right] + \left\|\mathsf{Bias}\left(\hat{\mathbf{v}}_{\mathrm{ML}}^{\mathrm{atk}}\right)\right\|_2^2.
\label{eq:MSEbias}
\end{equation}
\end{theorem}
\begin{IEEEproof}
By adding and subtracting $\bm\mu_{\mathrm{ML}}^{\mathrm{atk}}$ (cf. Theorem~\ref{prop:MLstats}) from the term $(\hat{\mathbf{v}} - \mathbf{v})$ we obtain:
\begin{align}
&\mathsf{MSE}\left(\hat{\mathbf{v}}_{\mathrm{ML}}^{\mathrm{atk}}\right) = \notag \\
& E\left[(\hat{\mathbf{v}} - \bm\mu_{\mathrm{ML}}^{\mathrm{atk}} + \bm\mu_{\mathrm{ML}}^{\mathrm{atk}} - \mathbf{v})^\top
(\hat{\mathbf{v}} - \bm\mu_{\mathrm{ML}}^{\mathrm{atk}} + \bm\mu_{\mathrm{ML}}^{\mathrm{atk}} - \mathbf{v})\right] = \notag \\
& E\left[(\hat{\mathbf{v}} - \bm\mu_{\mathrm{ML}}^{\mathrm{atk}})^\top (\hat{\mathbf{v}} - \bm\mu_{\mathrm{ML}}^{\mathrm{atk}})\right] +
(\bm\mu_{\mathrm{ML}}^{\mathrm{atk}} - \mathbf{v})^\top
(\bm\mu_{\mathrm{ML}}^{\mathrm{atk}} - \mathbf{v}) \notag \\
& \mspace{140mu} + \: 2 E\left[(\hat{\mathbf{v}} - \bm\mu_{\mathrm{ML}}^{\mathrm{atk}})^\top (\bm\mu_{\mathrm{ML}}^{\mathrm{atk}} - \mathbf{v})\right]. \label{eq:pf-MSEbias}
\end{align}
The first term in~\eqref{eq:pf-MSEbias} is the trace of the covariance matrix $\mathrm{cov}[\hat{\mathbf{v}}_{\mathrm{ML}}^{\mathrm{atk}}]$ (cf. Theorem~\ref{prop:MLstats}). The second term is equal to $\left\|\mathsf{Bias}\left(\hat{\mathbf{v}}_{\mathrm{ML}}^{\mathrm{atk}}\right)\right\|_2^2$. Expanding the third term reveals that it is zero. Therefore,~\eqref{eq:MSEbias} follows.  
\end{IEEEproof}
The theorem states that as the norm of the bias increases, the MSE increases and vice versa. Recall that  $\mathbf{G}$ depends only on the network topology and the location of the installed PMUs.

Power system protection, control, and status verification are all contingent upon the availability of accurate state estimates. For example, voltage stability assessment (VSA) computes the load change that a system can tolerate before voltage collapse occurs. To perform VSA, the current operating point is obtained from the state estimator~\cite{Ejebe-VSA}, and thus corrupted state estimates can have significant impact on the power grid security. It is worth emphasizing that even if a very small number of PMUs are attacked, the estimated voltages at all buses are affected.  By leveraging the previously defined SE accuracy metrics, the next section studies how the location and angles of the attacked PMUs affect the quality of the state estimates. 

\begin{remark}
Apart from the MSE and bias, which are general metrics of the SE accuracy, additional indices are used for evaluating the severity of contingencies~\cite{HazraCatastrophic} based on the current network state. Since GPS spoofing disturbs the SE, it also affects the computation of contingency severity indices. The methodology in Sections~\ref{subsec:vulnerable} and~\ref{ssec:sol_approach} can also be applied to these indices.
\end{remark}

\subsection{The Most Vulnerable PMU Location}\label{subsec:vulnerable}

The objective of this section is to furnish the system operator with an analytical tool to study how the SE accuracy is affected by different attack combinations, and identify the PMUs that can induce the largest bias or MSE if attacked. Theorem~\ref{thm:MSE} reveals that the MSE and the squared norm of the bias are related via an additive constant.  Therefore, any of the two metrics can be used to study the effects that the spoofing attacks have on the SE accuracy. An optimization problem  to find the attack angle combinations that maximize the norm of the bias is formulated in the sequel.

Specifically, the optimization variables are $\mathbf{b} \in \{0,1\}^{N_b}$, which denotes the vector of attacked PMUs, and the attack angles $\bm{\Delta\theta}$. The optimization problem to maximize the bias, \emph{solved by the network operator,} is stated next, where $N_p$ be the number of PMUs attacked and an upper bound $\bm{\Delta\theta}_{\max}$ on the attack angle is considered to potentially account for the attacker's limited capability to shift the phasor angle:\footnote{For vectors $\mathbf{x},\mathbf{y}\in\mathbb{R}^N$, notation $\mathbf{x} \preceq \mathbf{y}$ means $x_i\leq y_i$, $i=1,\ldots,N$.}
\begin{subequations}\label{eq:Add1_eq}
\begin{align}
& \underset{\mathbf{\Delta \theta},\ \mathbf{b}}{\text{maximize}}
	& & ||\mathbf{B}_{\mathrm{ML}} (\bm{\Delta\theta}) \mathbf{v}||_2 \label{eq:Add1_eq:a}  \\               	
		& \text{subject to}
		& & -\bm{\Delta\theta}_{\max} \odot \mathbf{b} \preceq \bm{\Delta\theta} \preceq \bm{\Delta\theta}_{\max} \odot \mathbf{b},\label{eq:Add1_eq:b} \\
		&&& \mathbf{b} \in \left\{0,1\right\}^{N_b}, \; \mathbf{b} \preceq \mathbf{a}, \label{eq:Add1_eq:c}\\
		&&& \sum_{i=1}^{N_b}{b_i} = N_p \label{eq:Add1_eq:e}
\end{align}
\end{subequations}
where $\odot$ represents entry-wise multiplication.

The two optimization variables in \eqref{eq:Add1_eq} are $\bm{\Delta\theta}$ and $\mathbf{b}$. Constraint \eqref{eq:Add1_eq:b} imposes bounds on $\bm{\Delta\theta}$ that account for the maximum phase shift induced by the attack. The bounds $\bm{\Delta\theta}_{\max}$ can be set empirically based on studies such as the ones listed in Table~\ref{table:1}. Constraint~\eqref{eq:Add1_eq:c} expresses the binary nature of $\mathbf{b}$ and represents that an attack can only happen on buses where PMUs are installed. The number of PMUs attacked in the network is captured via constraint \eqref{eq:Add1_eq:e}. The value of $N_p$ is realistically set to a small number, because it is unlikely that multiple GPS spoofers simultaneously attack at different geographical locations.

Problem \eqref{eq:Add1_eq} is a mixed integer program due to the binary $\mathbf{b}$ and thus hard to solve. Furthermore, even if the binary variables are relaxed to intervals $0 \leq b_i \leq 1$, the resulting problem is still nonconvex due to presence of sinusoids in the objective function.
 
Optimization problem~\eqref{eq:Add1_eq} yields the bus locations that if attacked induce the largest bias. The inputs are $\mathbf{a}$, $\mathbf{v}$, $N_p$, and  $\bm{\Delta\theta}_{\max}$. The vector $\mathbf{a}$ represents the buses with installed PMUs. It is worth emphasizing that optimization problem~\eqref{eq:Add1_eq} is solved by the system operator, with the purpose of identifying the attacks that can potentially cause the maximum deviation of the state estimate from its true value. 
Optimization problem~\eqref{eq:Add1_eq}  depends on the voltage profile $\mathbf{v}$. The voltage profile is typically determined as a result of an optimal power flow routine, which considers the network demand and optimizes economical objectives, subject to reliability constraints (such as line thermal limits). The system operator can thus solve~\eqref{eq:Add1_eq} for different voltage profiles $\mathbf{v}$, as the nodal injections and loads vary. Section~\ref{ssec:sol_approach} details the solution approach for~\eqref{eq:Add1_eq}.

\subsection{Solution Approach} \label{ssec:sol_approach}

Problem \eqref{eq:Add1_eq} can be optimally solved by enumerating all possible combinations of vector $\mathbf{b}$ satisfying constraints~\eqref{eq:Add1_eq:c} and \eqref{eq:Add1_eq:e}, and solving for each vector $\mathbf{b}$ the following optimization problem with variable $\bm{\Delta\theta}$:
\begin{equation}\label{eq:Add2}
	\begin{aligned}
		& \underset{\bm{\Delta\theta}}{ \text{maximize}}
		& & ||\mathbf{B}_{\mathrm{ML}} (\bm{\Delta\theta}) \mathbf{v}||_2 \\
		& \text{subject to}
		& & -\bm{\Delta\theta}_{\max} \odot \mathbf{b} \preceq \bm{\Delta\theta} \preceq \bm{\Delta\theta}_{\max} \odot \mathbf{b} \\
	\end{aligned}
\end{equation}
Recall that the bias $\mathbf{B}_{\mathrm{ML}} (\bm{\Delta\theta}) \mathbf{v}$ is a nonconvex function of $\bm{\Delta\theta}$.  We use MATLAB's  nonlinear programming solver \texttt{fmincon} \cite{fmincon} to find a  local maximum of~\eqref{eq:Add2}. As the solution depends on the initial point, it is better to solve \eqref{eq:Add2} for multiple initializations. The solution of~\eqref{eq:Add1_eq} is given by $\mathbf{b}$ and $\bm{\Delta\theta}$ resulting in the largest objective value for~\eqref{eq:Add2}. The optimal procedure to solve~\eqref{eq:Add1_eq} is given in Algorithm~\ref{alg:Optimal}.

\begin{algorithm}[!t]
	\SetAlgoLined 
	\caption{\textbf{Optimal Algorithm} to identify the $N_p$ most vulnerable PMU locations}
	\label{alg:Optimal}
	\textbf{Input}: $ \mathbf{a}, \mathbf{v}, \bm{\Delta\theta}_{\text{max}}, \text{and}\ N_p$
	
	  Obtain the combinations  $c=\left(\begin{smallmatrix} |\mathcal{N}_{\mathrm{PMU}}| \\ N_p \end{smallmatrix}\right)$---each combination corresponds to a specific $\mathbf{b}$
	  
	\For {$i = 1:c$} {
				Solve \eqref{eq:Add2} using three initializations: $\bm{\Delta\theta}_{\mathrm{init}} = \bm{0}$, 
$\bm{\Delta\theta}_{\mathrm{init}} = -\bm{\Delta\theta}_{\mathrm{max}} \odot  \mathbf{b}$, 
$\bm{\Delta\theta}_{\mathrm{init}} =  \bm{\Delta\theta}_{\mathrm{max}} \odot  \mathbf{b} $

				Record the largest objective value resulting from the three initializations and the corresponding $\bm{\Delta\theta}$
				     }
	Find the largest among the $c$ recorded values.  The corresponding vectors $\mathbf{b}$ and $\bm{\Delta\theta}$ are the solution to \eqref{eq:Add1_eq}.
\end{algorithm}
   
\begin{algorithm}[!t]
	\SetAlgoLined
	\caption{\textbf{Greedy Algorithm} to identify $N_p=2$ most vulnerable PMU locations}
	\label{alg:Greedy}
	\textbf{Input}: $\mathbf{a}, \mathbf{v}, \bm{\Delta\theta}_{\mathrm{max}}$; worst PMU location and attack ($n_1$, $\Delta\theta_{n_1}^{*}$)  from Algorithm \ref{alg:Optimal} 
		  with $N_p=1$

    \For {$i = 1:N_b$}
    				{
    \If {$\text{a}_{i} = 1 \ \ \text{and}\ \ i \neq n_1$}
    					{
							$b_i = 1$, $b_{n_1} = 1$, $b_j = 0 \  \forall\ j \neq \{i,n_1\}$; \\ 
							Solve \eqref{eq:Add2} with constraint $\Delta\theta_{n_1} = \Delta\theta_{n_1}^{*}$ and  initializations $\Delta\theta_i \in \{0,-{\Delta\theta}_{i\,\mathrm{max}},  {\Delta\theta}_{i\,\mathrm{max}} \}$

Choose $\bm{\Delta\theta}^{*}$ corresponding to the largest objective value and obtain $||\mathbf{B}_{\mathrm{ML}} (\bm{\Delta\theta}^{*}) \mathbf{v}||_2$\\

						Reset $b_i = 0$
           			    }
           			 }
     The second most vulnerable PMU location corresponds to the largest recorded $||\mathbf{B}_{\mathrm{ML}} (\bm{\Delta\theta}^*) \mathbf{v}||_2$
\end{algorithm}

Algorithm \ref{alg:Optimal} uses exhaustive search. It thus has substantial complexity for  large networks especially when $N_p\geq 2$, because it considers all combinations of $N_p$ PMUs. An alternative approach is to use a greedy algorithm. 
Consider the case of $N_p=2$. The main idea is to use Algorithm~\ref{alg:Optimal} with $N_p=1$ to identify the single most vulnerable PMU and corresponding attack angle, then fix the attack angle on that bus and search for a second PMU bus and attack angle.  

Specifically, first Algorithm \ref{alg:Optimal} is run with $N_p=1$.  Let $n_1$ be the resulting  bus index, that is $b_{n_1}^{*}=1$, and let $\Delta\theta_{n_1}^{*}$ be the corresponding optimal attack angle.  The second step is to solve problem~\eqref{eq:Add1_eq} with $N_p=2$ and additional constraints $b_{n_1}=1$ and $\Delta\theta_{n_1}=\Delta\theta_{n_1}^{*}$. This approach effectively reduces the problem of considering all combinations of $N_p=2$ PMUs into two searches of $N_p=1$ PMU, which are much simpler. Algorithm \ref{alg:Greedy} summarizes the steps of the greedy approach. The idea can be extended to larger values of $N_p$. For example, if $N_p=3$, Algorithm~\ref{alg:Greedy} can be applied to find the first two most vulnerable PMUs (denote them as $n_1$  and $n_2$ with corresponding angles $\Delta\theta_{n_1}^{*}$ and $\Delta\theta_{n_2}^{*}$). Then, problem~\eqref{eq:Add1_eq} with $N_p=3$ and additional constraints $b_{n_1}=1$, $b_{n_2}=1$, $\Delta\theta_{n_1}=\Delta\theta_{n_1}^{*}$,  and $\Delta\theta_{n_2}=\Delta\theta_{n_2}^{*}$ is solved. 

The chief reason why Algorithm~\ref{alg:Greedy} has potential for computational effectiveness relative to Algorithm~\ref{alg:Optimal} is the complexity of the nonconvex optimization problem~\eqref{eq:Add2} that needs to be solved in every iteration of these algorithms.
In particular, problem~\eqref{eq:Add2} always has \emph{one} optimization variable for all cases that needs to be solved as Algorithm~\ref{alg:Greedy}  runs. In contrast, problem~\eqref{eq:Add2} always has $N_p$ optimization variables for all cases that needs to be solved as  Algorithm~\ref{alg:Optimal} runs. Taking advantage of the fact that highly efficient algorithms exist for one-dimensional minimization, the reduction in the number of optimization variables plays a significant role.

\section{State Estimation and Attack Reconstrucion}\label{sec:AM}
This section develops an algorithm to jointly estimate the system state and identify the phase shifts for attacked buses. 

The network operator has  access to the measurement vectors $\mathbf{z}_n^{\text{atk}}$ at all buses on which PMUs are installed. 
{ Using the model in \eqref{eq:znmeas}, the problem of jointly estimating $\mathbf{v}$ and ${\Delta \theta}_{n}$ amounts to the following nonlinear least squares problem with variables  $\mathbf{v}$ and  $\{\Delta\theta_n\}_{n \in \mathcal{N}_\text{PMU}}$, where for clarity, the dependency of $\bm{\Gamma}_n$ on $\Delta\theta_n$ is denoted explicitly:
\begin{equation}
\label{eq:NLS1}
\min_{\mathbf{v}, \bm{\Delta\theta}} 
 \sum_{n=1}^{N_b}a_n(\mathbf{z}_n^{\text{atk}} - \bm{\Gamma}_n(\Delta\theta_n) \mathbf{H}_{n}\mathbf{v})^{\top}\bm{\Sigma}_n^{-1}(\mathbf{z}_n^{\text{atk}} - \bm{\Gamma}_n(\Delta\theta_n) \mathbf{H}_{n}\mathbf{v}). 
\end{equation}

The previous problem is nonlinear due to the trigonometric functions present in the definition of $\bm{\Gamma}_n(\Delta\theta_n)$ (cf.~Section~\ref{subsec:TSAmodel}). To alleviate this nonlinearity, we change the variable $\Delta\theta_n$ by introducing a new optimization variable $\bm{\gamma}_n =  \left[ \begin{smallmatrix} \gamma_{n,1} \\ \gamma_{n,2} \end{smallmatrix} \right]=  \left[ \begin{smallmatrix} \cos(\Delta\theta_{n}) \\ \sin(\Delta\theta_{n}) \end{smallmatrix} \right]$. In order to be able to uniquely recover $\Delta\theta_n$, the constraint $\gamma_{n,1}^2 + \gamma_{n,2}^2=1$ is added. The resulting problem is a constrained bilinear least squares problem with optimization variables $\mathbf{v}$ and  $\{\bm{\gamma}_n\}_{n \in \mathcal{N}_\text{PMU}}$:
\begin{subequations}
\label{eq:AM}  
\begin{align}  
		 \underset{\mathbf{v}, \{\bm{\gamma}_n\}_{n \in \mathcal{N}_{\mathrm{PMU}}}}{\text{minimize}}~~~
		&  \sum_{n=1}^{N_b} f(\bm{\gamma}_n, \mathbf{v}) \label{eq:AM_obj}  \\
		\text{subject to~~~~~~}
		&  \bm{\gamma}_n^{\top} \bm{\gamma}_n = 1, \ \ n \in \mathcal{N}_{\mathrm{PMU}} , \label{eq:AM_constr}
\end{align}
\end{subequations}
where the objective function $f(\bm{\gamma}_n, \mathbf{v})$ 
is written as
$$ f(\bm{\gamma}_n, \mathbf{v}) = a_n(\mathbf{z}_n^{\text{atk}} - \bm{\Gamma}_n \mathbf{H}_{n}\mathbf{v})^{\top}\bm{\Sigma}_n^{-1}(\mathbf{z}_n^{\text{atk}} - \bm{\Gamma}_n \mathbf{H}_{n}\mathbf{v}) $$ 
and $\bm{\Gamma}_n$ is a block diagonal matrix that includes $2\times 2$ blocks of the term $\left[ \begin{smallmatrix} \gamma_{n,1} & -\gamma_{n,2}  \\ \gamma_{n,2} & \gamma_{n,1} \end{smallmatrix} \right]$; the variables $\bm{\gamma}_n$ and $\bm{\Gamma}_n$ are used interchangeably.  
}

Problem~\eqref{eq:AM} is nonconvex and thus challenging. The nonconvexity arises because the objective is bilinear in the variables $\mathbf{v}$ and $\bm{\gamma}_n$ and also because of the quadratic equality constraint~\eqref{eq:AM_constr}. But problem~\eqref{eq:AM} can be efficiently tackled via an alternating minimization (AM) algorithm as explained in the sequel.

In general, the AM algorithm is applicable to minimization with respect to two groups of variables, in this case $\mathbf{v}$ and $\{\bm{\gamma}_n\}_{n \in \mathcal{N}_\text{PMU}}$. In the first step, minimization with respect to the first group is performed, by assuming the second group  is kept fixed. The second step consists of minimization with respect to the second group of variables upon substituting the updated values for the first group of variables. These two steps are repeated until convergence.  
The convergence criterion is  $|$\texttt{CurrObj} $-$ \texttt{PrevObj}$| / |$\texttt{CurrObj}$|$ $\leq$ Tolerance  $(\epsilon)$,
where \texttt{PrevObj} and \texttt{CurrObj} represent the objective function values \eqref{eq:AM_obj} before and after the update. The algorithm is initialized by setting $\bm{\gamma}_n = \bmat{1 & 0}^{\top}$ for all $n\in\mathcal{N}_{\mathrm{PMU}}$.

The AM algorithm has the attractive feature that it decreases the objective~\eqref{eq:AM_obj}  in every iteration. Interestingly, the minimizations in this algorithm can be solved in closed form, as shown next.

\subsection{Minimization With Respect to the State} \label{sub:Min_v}
The minimization of~\eqref{eq:AM} with respect to $\mathbf{v}$ amounts to unconstrained least squares with solution 
\begin{align}\label{eq:AM_Sol}
\hat{\mathbf{v}}_{\mathrm{AM}} = \mathbf{G}_{\mathrm{AM}}^{-1} \sum_{n=1}^{N_b} a_n (\mathbf{\Gamma}_n \mathbf{H}_n)^{\top} \mathbf{\Sigma}_n^{-1} \mathbf{z}_n^{\text{atk}}
\end{align}
where it is assumed that the matrix $\mathbf{G}_{\mathrm{AM}}= \sum_{n=1}^{N_b}a_n(\mathbf{\Gamma}_n \mathbf{H}_n)^{\top}\bm{\Sigma}_n^{-1}(\mathbf{\Gamma}_n \mathbf{H}_n)$ is non-singular.

\subsection{Minimization With Respect to the Attack Angle} \label{sub:Min_gamma}
The minimization in \eqref{eq:AM} with respect to $\bm{\gamma}_n$ takes the following equivalent form: 
\begin{subequations}\label{eq:AM_Equiv}
	\begin{align}
		& \underset{\bm{\gamma}_n}{\text{minimize}}
		& & (\mathbf{z}_n^{\text{atk}} - \mathbf{A}_n \bm{\gamma}_n)^{\top}\bm{\Sigma}_n^{-1}(\mathbf{z}_n^{\text{atk}} - \mathbf{A}_n \bm{\gamma}_n)  \label{eq:AM_Equiv:a}\\
		& \text{subject to}
		& & \bm{\gamma}_n^{\top} \bm{\gamma}_n = 1 \label{eq:AM_Equiv:b}
	\end{align}
\end{subequations}
where $\mathbf{h}_{n,i}^{\top}$ is the $i$-th row of $\mathbf{H}_n \ (i=1,2,\dotsc, M_n)$ and $\mathbf{A}_{n} \in \mathds{R}^{M_n\times 2}$ is defined as
\begin{equation}
\mathbf{A}_n = 
\begin{bmatrix}
\mathbf{h}_{n,1}^{\top} \mathbf{v} & -\mathbf{h}_{n,2}^{\top} \mathbf{v} \\ \mathbf{h}_{n,2}^{\top} \mathbf{v} & \mathbf{h}_{n,1}^{\top} \mathbf{v} \\ \vdots & \vdots  \\ \mathbf{h}_{n,M_n-1}^{\top} \mathbf{v} & -\mathbf{h}_{n,M_n}^{\top} \mathbf{v} \\ \mathbf{h}_{n,M_n}^{\top} \mathbf{v} & \mathbf{h}_{n,M_n-1}^{\top} \mathbf{v}
\end{bmatrix}.
\end{equation}

Problem~\eqref{eq:AM_Equiv} is nonconvex due to the quadratic equality constraint~\eqref{eq:AM_Equiv:b}. Interestingly, it is possible to solve this problem in closed form. To facilitate the solution, consider the eigenvalue decomposition (EVD) of  $\mathbf{A}_{n}^{\top} \mathbf{\Sigma}_n^{-1} \mathbf{A}_{n}$
given as $\mathbf{A}_{n}^{\top} \mathbf{\Sigma}_n^{-1} \mathbf{A}_{n} = \mathbf{Q} \mathbf{\Xi} \mathbf{Q}^{\top}$,
where $\mathbf{Q} \in \mathds{R}^{2 \times 2}$ is orthonormal and $\mathbf{\Xi} \in \mathds{R}^{2 \times 2}$ is a diagonal matrix of non-negative eigenvalues $\xi_1$ and $\xi_2$. Define further $\mathbf{u}_n = \mathbf{Q}^{\top} \mathbf{A}_{n}^{\top} \mathbf{\Sigma}_n^{-1} \mathbf{z}_n^{\text{atk}}  = \bmat{u_{n,1} & u_{n,2}}^{\top}$. Notice that $\mathbf{Q}$ and $\mathbf{u}_n$ are readily obtained with knowledge of the current iterate $\mathbf{v}$ and the measurement $\mathbf{z}_n^{\text{atk}}$. The next theorem characterizes the solution to \eqref{eq:AM_Equiv}.
\begin{theorem}\label{thm:gamma}
The minimizer of \eqref{eq:AM_Equiv} is given by
\begin{equation}\label{eq:gamma}
  \bm{\gamma}_n = (\mathbf{A}_{n}^{\top} \mathbf{\Sigma}_n^{-1} \mathbf{A}_{n} + \lambda_n \mathbf{I})^{-1} \mathbf{A}_{n}^{\top} \mathbf{\Sigma}_n^{-1} \mathbf{z}_n^{\text{atk}},
\end{equation}
where $\lambda_n$ is the Lagrange multiplier corresponding to \eqref{eq:AM_Equiv:b} and is a root of the following quartic equation in $\lambda_n$ which has at least one real solution:
	\begin{equation}\label{eq:decompose_3}
		g(\lambda_n) = \frac{u_{n,1}^{2}}{(\xi_1+\lambda_n)^2} + \frac{u_{n,2}^{2}}{(\xi_2+\lambda_n)^2} = 1
	\end{equation}	
\end{theorem}
\begin{IEEEproof}
An optimal Lagrange multiplier always exists for~\eqref{eq:AM_Equiv}, as the linear independence constraint qualification holds~\cite[Sec.~3.1]{BertsekasNLP2nd}.
The Lagrangian function of \eqref{eq:AM_Equiv} is
\begin{align}\label{eq:AM_L}
L(\bm{\gamma}_n,\lambda_n) &= (\mathbf{z}_n^{\text{atk}} - \mathbf{A}_n \bm{\gamma}_n)^{\top}\bm{\Sigma}_n^{-1}(\mathbf{z}_n^{\text{atk}} - \mathbf{A}_n \bm{\gamma}_n) + \notag\\
 &+ \lambda_n (\bm{\gamma}_n^{\top} \bm{\gamma}_n - 1) .
\end{align}
The optimality condition that yields $\bm{\gamma}_n$ as a function of $\lambda_n$ is given by $ \nabla_{\bm{\gamma}_n} L(\bm{\gamma}_n,\lambda_n) = \bm 0 $, which yields
\begin{equation} \label{eq:Opt_cond}
(\mathbf{A}_{n}^{\top} \mathbf{\Sigma}_n^{-1} \mathbf{A}_{n} + \lambda_n \mathbf{I})\bm{\gamma}_n = \mathbf{A}_{n}^{\top} \mathbf{\Sigma}_n^{-1} \mathbf{z}_n^{\text{atk}}.
\end{equation}
Assuming the invertibility of $(\mathbf{A}_{n}^{\top} \mathbf{\Sigma}_n^{-1} \mathbf{A}_{n} + \lambda_n \mathbf{I})$---which will be discussed shortly---\eqref{eq:gamma} is obtained.

To find $\lambda_n$, substitute \eqref{eq:gamma} into the constraint $\bm{\gamma}_n^{\top} \bm{\gamma}_n = 1$. This yields the equation $g(\lambda_n)=1$
where
\begin{gather*}
g(\lambda_n) = {\mathbf{z}_n^{\text{atk}}}^{\top} \mathbf{\Sigma}_n^{-1} \mathbf{A}_n (\mathbf{A}_{n}^{\top} \mathbf{\Sigma}_n^{-1} \mathbf{A}_{n} + \lambda_n \mathbf{I})^{-2} \mathbf{A}_n^{\top} \mathbf{\Sigma}_n^{-1} \mathbf{z}_n^{\text{atk}}.
\end{gather*}

The function $g(\lambda_n)$ can be written in the following simpler form if $\mathbf{A}_{n}^{\top} \mathbf{\Sigma}_n^{-1} \mathbf{A}_{n}$ is substituted by its EVD $\mathbf{Q} \mathbf{\Xi} \mathbf{Q}^{\top}$:
\begin{equation}\label{eq:decompose_2}
g(\lambda_n) = \mathbf{u}_n^{\top} (\mathbf{\Xi} + \lambda_n \mathbf{I})^{-2} \mathbf{u}_n
\end{equation}
Eq. \eqref{eq:decompose_3} is obtained by expanding \eqref{eq:decompose_2}.

The function $g(\lambda_n)$ is strictly decreasing in $\lambda_n$ in interval $[-\min\{\xi_1,\xi_2\},\infty)$. Moreover, the limit of $g(\lambda_n)$ as $\lambda_n$ approaches $-\min\{\xi_1,\xi_2\}$ or $+\infty$ is  respectively $+\infty$ or $0$, therefore, the equation $g(\lambda_n)=1$ has one real root in the interval $[-\min\{\xi_1,\xi_2\},\infty)$.
\end{IEEEproof}

Eq. \eqref{eq:decompose_3} is quartic in $\lambda_n$, that is, it yields four values of $\lambda_n$ as solutions. The solutions of a quartic equation are completely characterized and can be routinely computed; see e.g.,~\cite{Quartic}. In addition, Theorem~\ref{thm:gamma} ensures that at least one real solution exists, i.e., a situation where~\eqref{eq:decompose_3} has only complex roots never arises. The optimal value of $\lambda_n$ is obtained by substituting each value of $\lambda_n$ first in \eqref{eq:gamma}, and then picking the one that yields the smallest objective in \eqref{eq:AM_Equiv}. Algorithm \ref{alg:AM Algorithm} summarizes the steps to solve~\eqref{eq:AM}.

The previous procedure works seamlessly, unless $\mathbf{A}_{n}^{\top} \mathbf{\Sigma}_n^{-1} \mathbf{A}_{n} + \lambda_n \mathbf{I}$ is not invertible, which is unlikely in practice, as explained next. In particular, the following theorem characterizes the invertibility of $\mathbf{A}_{n}^{\top} \mathbf{\Sigma}_n^{-1} \mathbf{A}_{n} + \lambda_n \mathbf{I}$.
\begin{theorem}
The matrix $\mathbf{A}_{n}^{\top} \mathbf{\Sigma}_n^{-1} \mathbf{A}_{n} + \lambda_n \mathbf{I}$ in \eqref{eq:gamma} is invertible if ${u}_{n,1} \neq 0$ or ${u}_{n,2} \neq 0$.
\end{theorem}

\begin{IEEEproof} We will prove the claim by contradiction. Suppose that the matrix $\mathbf{A}_{n}^{\top} \mathbf{\Sigma}_n^{-1} \mathbf{A}_{n} + \lambda_n \mathbf{I}$ is not invertible. Then, the optimal Lagrange multiplier $\lambda_n$ is $-\xi_1$ or $-\xi_2$.

Using the EVD of $\mathbf{A}_{n}^{\top} \mathbf{\Sigma}_n^{-1} \mathbf{A}_{n}$, \eqref{eq:Opt_cond} is written as
\begin{equation}\label{eq:EVD}
(\mathbf{\Xi} + \lambda_n \mathbf{I}) \mathbf{Q}^{\top} \bm{\gamma}_n = \mathbf{u}_n
\end{equation}
If  $\lambda_n = -\xi_1$,~\eqref{eq:EVD} takes the form
$\left[ \begin{smallmatrix} 0 & 0 \\ 0 & \xi_2-\xi_1  \end{smallmatrix} \right] \mathbf{Q}^{\top} \bm{\gamma}_n = \left[ \begin{smallmatrix} u_{n,1} \\ u_{n,2} \end{smallmatrix} \right]$. 
If $u_{n,1} \neq 0$, then the latter system of equations is incompatible. Thus, if  $u_{n,1} \neq 0$, the optimal multiplier cannot be $-\xi_1$.  A similar argument can be followed for $\lambda_n = -\xi_2$, which yields $\left[ \begin{smallmatrix} \xi_1-\xi_2 & 0 \\ 0 & 0  \end{smallmatrix} \right] \mathbf{Q}^{\top} \bm{\gamma}_n = \left[ \begin{smallmatrix} u_{n,1} \\ u_{n,2} \end{smallmatrix} \right]$.  
\end{IEEEproof}

In practice, it is unlikely that ${u}_{n,1} = 0$ or ${u}_{n,2} = 0$. The reason is that $\mathbf{u}_n$ depends on the noisy $\mathbf{z}_n^{\text{atk}}$ (recall that $\mathbf{u}_n= \mathbf{Q}^{\top} \mathbf{A}_{n}^{\top} \mathbf{\Sigma}_n^{-1} \mathbf{z}_n^{\text{atk}}$), and thus the noise must have very particular values in order to yield ${u}_{n,1} = 0$ or ${u}_{n,2} = 0$. In our numerical tests, we did not encounter any noninvertibility issue.

\begin{remark}[Relationship to works on imperfect synchronization]
The modeling of GPS spoofing is related to imperfect PMU synchronization~\cite{Yang}, and more broadly to asynchronous sampling in networked sensing systems~\cite{wai2015consensus}.  But the GPS spoofing attack is different than imperfect PMU synchronization. In the latter, the measurement delay  typically translates to less than $1^{\circ}$ for $60$ Hz frequency; whereas, in GPS spoofing, the attacker can potentially cause a much larger phase shift as documented in Table \ref{table:Phase}.  The small phase shift implies that approximations $\sin\Delta\theta_n\approx \Delta\theta_n$ and $\cos\Delta\theta_n\approx 1$ can be invoked to render $\bm\Gamma_n(\Delta\theta_n)$ linear. These approximations are used in~\cite{Yang}. Problem~\eqref{eq:NLS1} is more complicated when the attack angles are not assumed small. This section developed tractable algorithms for its solution based on the reformulation~\eqref{eq:AM} that includes the nonconvex constraint~\eqref{eq:AM_constr}. 

The work in~\cite{wai2015consensus} includes a general asynchronicity model, but relies upon computing the Fourier transform of the underlying signal to be estimated. As such, it is necessary to acquire a number of signal samples across time.  On the other hand, a single set of measurements from all PMUs is sufficient for the present work. The work~\cite{wai2015consensus} also develops an alternating minimization algorithm. It performs full optimization with respect to one set of variables, but takes one step of gradient descent with respect to the other set. On the other hand, the present work performs full minimizations with respect to each set of optimization variables, which becomes possible by exploiting Lagrangian duality and the structure of the particular problem at hand. 
\end{remark}

\subsection{Simplifications Under Diagonal Covariance}\label{subsec:diag}
Suppose that the covariance $\mathbf{\Sigma}_n$ is a diagonal matrix with equal variance for the real and imaginary parts of the voltage and likewise for the real and imaginary parts of every current on the $L_n$ lines \cite{Kekatos, Ali_Abur, Yang}.
The covariance is thus assumed to have  diagonal entries $\sigma_{n,i}^2$, $i=1,\ldots,M_n$,
where 
\begin{equation}
\sigma_{n,1} = \sigma_{n,2}, \: \sigma_{n,3} = \sigma_{n,4}, \: \dotsc, \: \sigma_{n,M_n-1}= \sigma_{n,M_n}.
\label{eq:covspcase}
\end{equation}
The computation of $\gamma_n$ is very simple in this case, as the following theorem describes.
\begin{theorem}
	The closed-form solution for $\bm{\gamma}_n$ when the covariance $\mathbf{\Sigma}_n$ has the structure given by~\eqref{eq:covspcase} is 
	\begin{equation}\label{eq:closed_form_3}
	\bm{\gamma}_n =  (1/ ||\mathbf{A}_n^{\top} \mathbf{\Sigma}_n^{-1} \mathbf{z}_n^{\text{atk}} ||_2)
	\mathbf{A}_n^{\top} \mathbf{\Sigma}_n^{-1} \mathbf{z}_n^{\text{atk}}.
	\end{equation} 
\end{theorem}

\begin{IEEEproof}
Under~\eqref{eq:covspcase}, the  matrix $\mathbf{A}_{n}^{\top} \mathbf{\Sigma}_n^{-1} \mathbf{A}_{n}$ is written as $\mathbf{A}_{n}^{\top} \mathbf{\Sigma}_n^{-1} \mathbf{A}_{n} = d_n \mathbf{I}$ 
where $d_n$ is given by 
\begin{align}\label{eq:compute_1}
d_n = \sum_{i=1,\; i\; \text{odd}}^{{M_n}-1} 
\sigma_{n,i}^{-2} 
[(\mathbf{h}_{n,i}^{\top} \mathbf{v})^{2} + (\mathbf{h}_{n,i+1}^{\top} \mathbf{v})^{2}].
\end{align}

Substituting $\mathbf{A}_{n}^{\top} \mathbf{\Sigma}_n^{-1} \mathbf{A}_{n} = d_n \mathbf{I}$  in $g(\lambda_n)=1$ yields
\begin{align*}
{\mathbf{z}_n^{\text{atk}}}^{\top} \mathbf{\Sigma}_n^{-1} \mathbf{A}_n (d_n \mathbf{I} + \lambda_n \mathbf{I})^{-2} \mathbf{A}_n^{\top} \mathbf{\Sigma}_n^{-1} \mathbf{z}_n^{\text{atk}} &= 1
\end{align*}
and the solution with respect to $\lambda_n$ is given by
$d_n + \lambda_n = \pm ||\mathbf{A}_n^{\top} \mathbf{\Sigma}_n^{-1} \mathbf{z}_n^{\text{atk}} ||_2$.
Substituting the latter into \eqref{eq:gamma}, we obtain two possible values of  $\bm{\gamma}_n$  given by 
$\bm{\gamma}_n(\lambda_n) = \pm {\mathbf{A}_n^{\top} \mathbf{\Sigma}_n^{-1} \mathbf{z}_n^{\text{atk}}}/{||\mathbf{A}_n^{\top} \mathbf{\Sigma}_n^{-1} \mathbf{z}_n^{\text{atk}} ||_2}$.
Substitution into \eqref{eq:AM_Equiv} reveals that only  the positive value of $\bm{\gamma}_n$ minimizes \eqref{eq:AM_Equiv}.
\end{IEEEproof}

The advantage of the diagonal covariance $\bm{\Sigma}_n$ is that there is no need to solve the quartic equation \eqref{eq:decompose_3} in order to obtain $\bm{\gamma}_n$, which simplifies the computation in Algorithm \ref{alg:AM Algorithm}. 
\begin{algorithm}[!t]
\SetAlgoLined
\caption{State Estimation \& Attack Reconstruction}
\label{alg:AM Algorithm}
\KwResult{ State Estimate $\hat{\mathbf{v}}_{\mathrm{AM}}$ and Attack Angle $\Delta\theta_n, n \in \mathcal{N}_{\text{PMU}}$ }
 \textbf{Input}: $\mathbf{z}_n^{\text{atk}}$
 
 Initialization: Solve \eqref{eq:AM_Sol} for $\hat{\mathbf{v}}_{\mathrm{AM}}$ by setting $\bm{\gamma}_n = [1\;0]^{\top}$
 
 \Repeat{convergence or maximum iterations reached}
  {
   \For {$n \in \mathcal{N}_{\text{PMU}}$}
    				{
  	Obtain 4 roots of $g(\lambda_n) = 1$ 
  	
  	Find the corresponding $\bm\gamma_n$ via \eqref{eq:gamma}
  	
  	Pick the $\bm{\gamma}_n$ that minimizes \eqref{eq:AM_Equiv:a}
  					}
  Update $\hat{\mathbf{v}}_{\mathrm{AM}}$ using \eqref{eq:AM_Sol}
 }
\end{algorithm}

The analyses of Sections \ref{sec:Attack_Formulation} and \ref{sec:AM} can be extended to SE with both PMU and SCADA measurements as explained next.

\section{Extensions to SE with PMU and SCADA measurements}\label{sec:PMU_SCADA}
PMUs have higher sampling rates than SCADA (0.008--0.03 sec compared to 2--4 sec)~\cite{Power}.
Consequently, combined SCADA and PMU measurements can increase the accuracy of SE~\cite[Sec.~III-C]{GGVKNG-spmag}.
State estimation with combined measurements is also referred to as hybrid SE. This section 
focuses on hybrid SE where SCADA measurements can be used as pseudo-measurements to provide a \textit{rough estimate} of the system state, which can then be used with PMU measurement for more accurate state estimation \cite{Phadake2008}. 
An overview of the hybrid SE is provided next, followed by the related analysis under GPS spoofing.

{
Due to the quadratic relation between the SCADA measurements and the state vector, the state vector can only be recovered up to a phase rotation~\cite{Kekatos}. To overcome this phase ambiguity issue in hybrid SE, as discussed in \cite{Kekatos}, a PMU is installed at reference bus, and furthermore, the phase of the reference bus is equivalently considered to be zero and removed from the state vector. In rectangular coordinates, $V_{1,i}$ is hence set to zero and removed from the state vector. Thus, the dimension of $\mathbf{v}$ is reduced to $(2N_b-1) \times 1$, and consequently, the $(N_b+1)$th column (row) is removed from the $\mathbf{H}_n\ (\mathbf{G})$ matrix in the linear model of Section~\ref{sec:se} involving PMU measurements.}

{
The SCADA measurement model is 
\begin{align}\label{SCADA_meas}
\mathbf{z}_s = \mathbf{h(\mathbf{v})} + \mathbf{w}_s
\end{align}
where $\mathbf{z}_s \in \mathbb{R}^{M}$ is the SCADA measurement vector, a nonlinear function $\mathbf{h}:\mathbb{R}^{2N_b-1} \rightarrow \mathbb{R}^{M}$ relates the state vector $\mathbf{v}$ to the measurement vector $\mathbf{z}_s$, and the noise vector is distributed according to $\mathbf{w}_s \sim \mathcal{N}(\mathbf{0}, \mathbf{\Sigma}_e)$, where $\mathbf{\Sigma}_e$ is a known positive definite covariance matrix of the SCADA measurements.  The SCADA state estimate $\hat{\mathbf{v}}_s$ is typically  obtained by solving the nonlinear least squares $\min_{\mathbf{v}} (\mathbf{z}_s - \mathbf{h(\mathbf{v})})^{\top} \mathbf{\Sigma}_e^{-1}  (\mathbf{z}_s - \mathbf{h(\mathbf{v})})$ via the Gauss-Newton method. Upon convergence, the covariance of $\hat{\mathbf{v}}_s$ can be approximated as follows~\cite{Kekatos}:
\begin{align}\label{eq:SCADA_cov}
	\mathbf{\Sigma}_s = \Big(\mathbf{J}^{\top}(\hat{\mathbf{v}}_s) \mathbf{\Sigma}_e^{-1} \mathbf{J}(\hat{\mathbf{v}}_s)\Big)^{-1}
\end{align}
where $\mathbf{J}(\mathbf{v}) = \nabla_{\mathbf{v}}\mathbf{h}(\mathbf{v})$ is the $M_n \times (2N_b-1)$ Jacobian matrix of $\mathbf{h}(\mathbf{v})$.
}

Hybrid SE can be cast in a Bayesian framework, where SCADA measurements are used to provide prior estimates $\hat{\mathbf{v}}_s$, which 
in turn are normally distributed with covariance $\mathbf{\Sigma}_s$ given previously.
The state estimate in the Bayesian framework is obtained from the maximum a posteriori (MAP) probability criterion:
 \begin{align} 
 	\hat{\mathbf{v}}_{\mathrm{MAP}} &= \underset{\mathbf{v}}{\mathrm{arg min}} \sum_{n=1}^{N_b}a_n(\mathbf{z}_n - \mathbf{H}_{n}\mathbf{v})^{\top}\bm{\Sigma}_n^{-1}(\mathbf{z}_n - \mathbf{H}_{n}\mathbf{v})  \notag \\
 	&\mspace{100mu}+ (\mathbf{v}-\hat{\mathbf{v}}_s)^{\top} \bm{\Sigma}_s^{-1} (\mathbf{v}-\hat{\mathbf{v}}_s).
 	\label{eq:mapcost}
 \end{align}
The last term in \eqref{eq:mapcost}  is a \textit{regularizer} that attracts the solution towards $\hat{\mathbf{v}}_s$, depending on how much the prior estimate is trusted, which is determined by $\bm\Sigma_s^{-1}$. The closed form solution $\hat{\mathbf{v}}_{\mathrm{MAP}}$ is derived similarly to $\hat{\mathbf{v}}_{\mathrm{ML}}$.
The MAP estimate is Gaussian distributed, that is, $\hat{\mathbf{v}}_{\mathrm{MAP}} \sim \mathcal{N}(\mathbf{v},\mathbf{G}_p^{-1})$, where $\mathbf{G}_p = \mathbf{G} + \bm{\Sigma}_s^{-1}$ is the regularized gain matrix and $\mathbf{G}$ is given previously. 

The state estimate using the MAP estimator with SCADA priors and corrupted measurement is given by
\begin{equation}
\label{eq:mapatk}
\hat{\mathbf{v}}_{\mathrm{MAP}}^{\mathrm{atk}} = \mathbf{G}_p^{-1}\left(\sum_{n=1}^{N_b}a_n\mathbf{H}_n^{\top}\bm{\Sigma}_n^{-1} \mathbf{z}_n^{\mathrm{atk}}+ \bm{\Sigma}_s^{-1}\hat{\mathbf{v}}_s\right).
\end{equation}
The statistics of \eqref{eq:mapatk} are given in \cite{Risbud2016}, where  it is observed that the attack introduces estimation bias. 

The analyses performed in Sections~\ref{sec:Attack_Formulation} and~\ref{sec:AM} can be easily extended to the case where both PMU and SCADA measurements are used for SE. In Section~\ref{sec:Attack_Formulation}, $\mathbf{B}_{\text{ML}}$ should be replaced by $\mathbf{B}_\text{MAP}$, given in \cite[eq. (23)]{Risbud2016}, to find the most vulnerable PMU in the network. { Similarly, an AM algorithm as in Section \ref{sec:AM} can be applied to combined PMU and SCADA measurements. In particular, the $\bm{\gamma}_n$ update for combined PMU and SCADA measurements is given by the solution to \eqref{eq:AM_Equiv}, because the regularizer in \eqref{eq:mapcost} does not contain $\bm{\gamma}_n$.  To update $\hat{\mathbf{v}}_\text{AM}$, \eqref{eq:AM_Sol} is replaced by the following equation:
\begin{multline}
\label{eq:AMmap}
\hat{\mathbf{v}}_{\mathrm{AM}} = (\mathbf{G}_{\mathrm{AM}} + \bm{\Sigma}_s^{-1})^{-1} \\ 
\cdot \left(\sum_{n=1}^{N_b}a_n(\mathbf{\Gamma}\mathbf{H}_n)^{\top}\bm{\Sigma}_n^{-1} \mathbf{z}_n^{\mathrm{atk}}+ \bm{\Sigma}_s^{-1}\hat{\mathbf{v}}_s\right).
\end{multline}
The alternating minimization steps are summarized in Algorithm~\ref{alg:AM Algorithm PMU and SCADA}. } 

\begin{algorithm}[!t]
	\SetAlgoLined
	\caption{State Estimation \& Attack Reconstruction for Combined PMU and SCADA Measurements}
	\label{alg:AM Algorithm PMU and SCADA}
	\KwResult{ State Estimate $\hat{\mathbf{v}}_{\mathrm{AM}}$ and Attack Angle $\Delta\theta_n, n \in \mathcal{N}_{\text{PMU}}$ }
	 \textbf{Inputs}: $\mathbf{z}_n^{\text{atk}}$ and SCADA-based estimate $\hat{\mathbf{v}}_s$
	 
	 Initialization:   Solve \eqref{eq:AMmap} for $\hat{\mathbf{v}}_{\mathrm{AM}}$ by setting $\bm{\gamma}_n = [1\;0]^{\top}$

	 \Repeat{convergence or maximum iterations reached}
	  {
	   \For {$n \in \mathcal{N}_{\text{PMU}}$}
	    				{
	  	Obtain 4 roots of $g(\lambda_n) = 1$ 
	  	
	  	Find the corresponding $\bm\gamma_n$ via \eqref{eq:gamma}
	  	
	  	Pick the $\bm{\gamma}_n$ that minimizes \eqref{eq:AM_Equiv:a}
	  					}
	  Update  $\hat{\mathbf{v}}_{\mathrm{AM}}$ using \eqref{eq:AMmap}
	 }
\end{algorithm}

\section{Numerical Tests}\label{sec:Num_tests}

This section presents numerical tests for identification of vulnerable PMU locations (Section~\ref{Worst_PMU}) as well as for the state and attack angle estimation using the AM algorithm (Sections~\ref{AM_detection} and~\ref{AM_SCADA_detection}). Sections~\ref{Bad_data} and~\ref{SpM} compare the performance of the AM algorithm with a bad data detector and with the work in~\cite{Fan2017}, respectively. All tests are performed on the IEEE 14-, 30-, and 118-bus networks.  All network parameters are provided in case files \texttt{case14.m}, \texttt{case30.m}, and \texttt{case118.m} of MATPOWER \cite{Zimmerman}, from which $\mathbf{H}_n$'s are computed.
The PMU placement vector $\mathbf{a}$ for all test cases is obtained using the criterion in \cite[eq. (7)]{Kekatos} via YALMIP \cite{YALMIP}, based exclusively on the availability of PMU measurements (i.e., setting $\mathbf{\Sigma}_s^{-1}=0$). Table \ref{table:1} lists the buses with installed PMUs for each network. The noise covariance $\mathbf{\Sigma}_n$ is diagonal resulting from standard deviation of 0.01 and 0.02 for bus voltage and line current measurements respectively. 

\begin{table}[!t]
\caption{Optimal PMU location ($\mathbf{a}$) for IEEE test networks.}
\label{table:1}
\centering
\begin{tabular}{ | m{4em} | m{2.5em}| m{5.5cm} | } 
\hline
 Test Case  & $|\mathcal{N}_{\text{PMU}}|$  & Bus number \\ 
\hline
 IEEE 14  & 6  & 2,4,6,7,10,14 \\ 
\hline
IEEE 30 & 13 & 2,3,6,10,11,12,15,20,23,25,27,28,29\\ 
\hline
\multirow{3}{4em}{IEEE 118} & \multirow{3}{2.5em}{94} & \multirow{6}{0cm}{}1--5,7--19,21--25,27--36,40,43,44,46,47,48,\\ 
& & 50,51,52,53,55--60,64,65,66,67,68,70,71,73,75,76,\\
& & 77,80--83,85--90,92,94--104,106--111,113--118\\ 
\hline
\end{tabular}
\end{table}

\subsection{Vulnerable PMU Location}\label{Worst_PMU}

We solve \eqref{eq:Add1_eq} for different network loads ranging from 50$\%$ to 150$\%$ of the nominal demand given in MATPOWER case files. 
Specifically, we calculate the corresponding voltage profile $\mathbf{v}$ resulting from 50$\%$ to 150$\%$ of the nominal real and reactive power demand using MATPOWER's \texttt{runpf}. Table~\ref{table:2} lists the most vulnerable PMU location for an attack on one PMU. 

\begin{table}[!t]
\caption{Most vulnerable PMU bus location for IEEE test networks using (case a)~nominal demand, (case b)~$50\%$ of nominal demand, and (case c)~$150\%$ of nominal demand.}
\label{table:2}
\centering
\begin{tabu} to 0.5\textwidth {  X[c] | X[c] | X[c] | X[c]}
\hline
 Test Case  & Bus [case (a)] & Bus [case (b)] & Bus [case (c)] \\
 \hline
 IEEE 14 & 6 & 6 & 6\\

 IEEE 30 & 12 & 12 & 12 \\

 IEEE 118  & 30 & 30 & 68 \\
\hline
\end{tabu}
\end{table}

\begin{figure*}[!t]
\centering
\subfloat[Voltage Magnitude Estimates]{\includegraphics[scale=0.21]{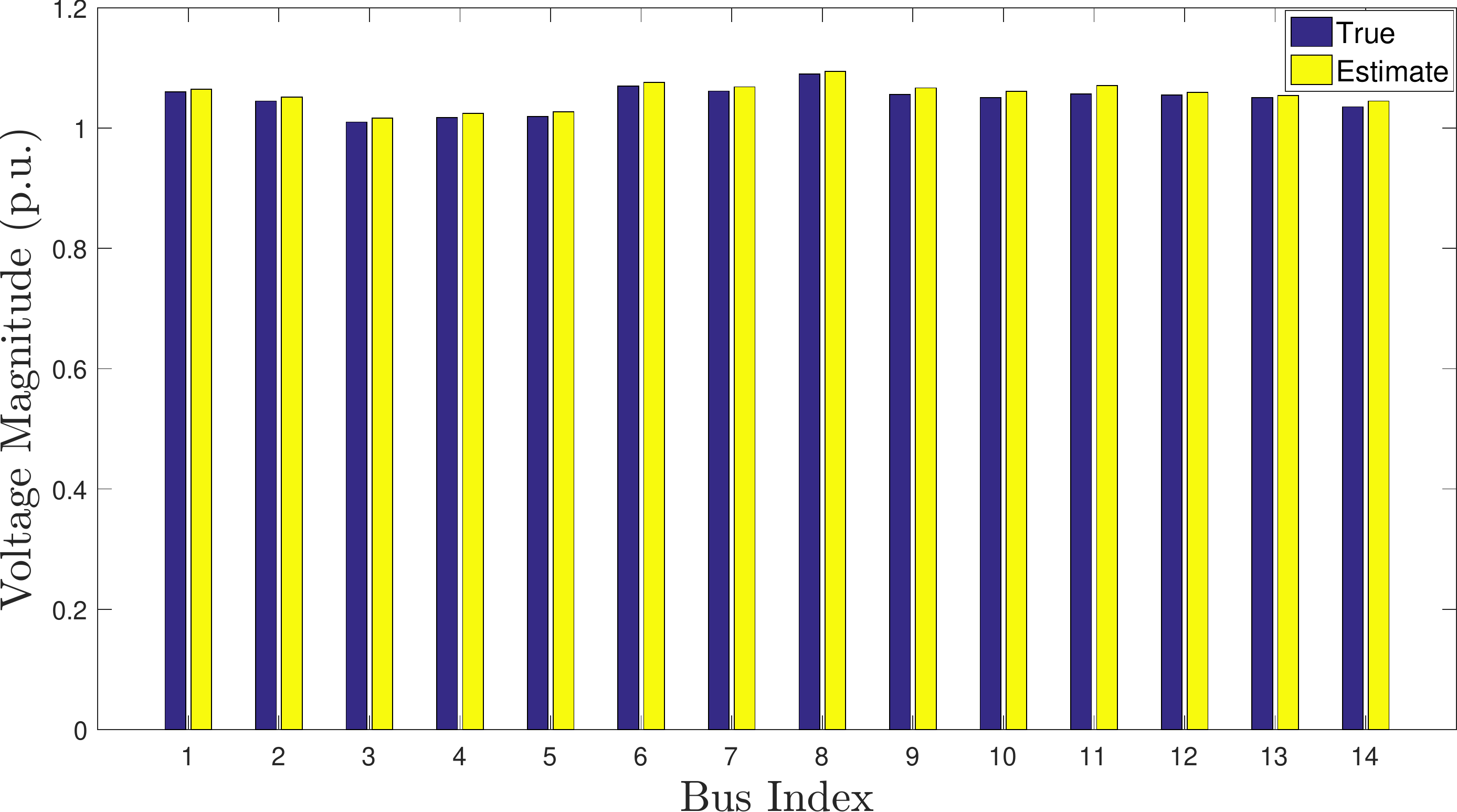}\label{fig:bar_magnitude}}
\hfil
\subfloat[Phase Angle Estimates]
{\includegraphics[scale=0.21]{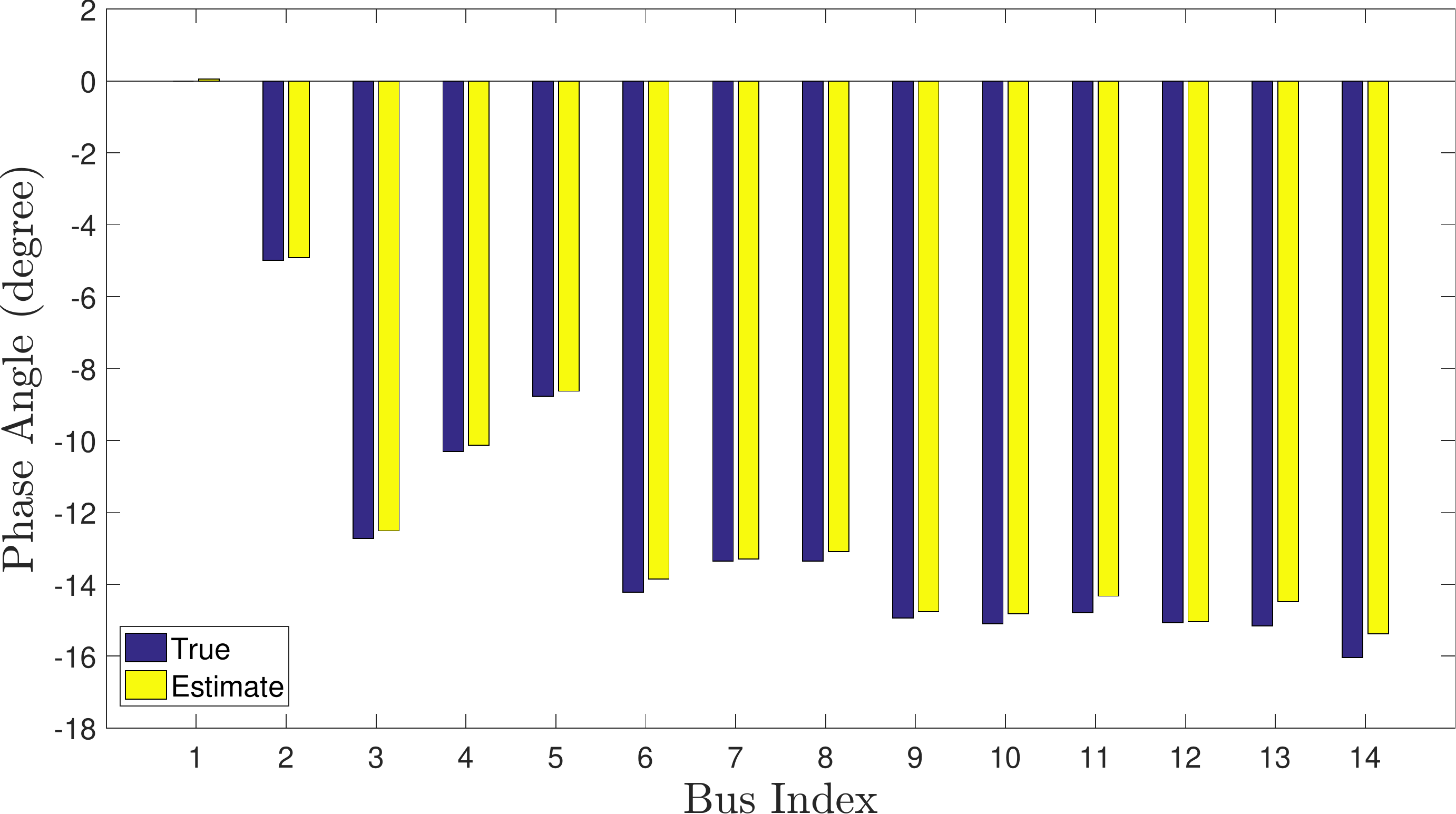}\label{fig:bar_phase}}
\hfil 
\subfloat[Attack Angle Estimates]
{\includegraphics[scale=0.21]{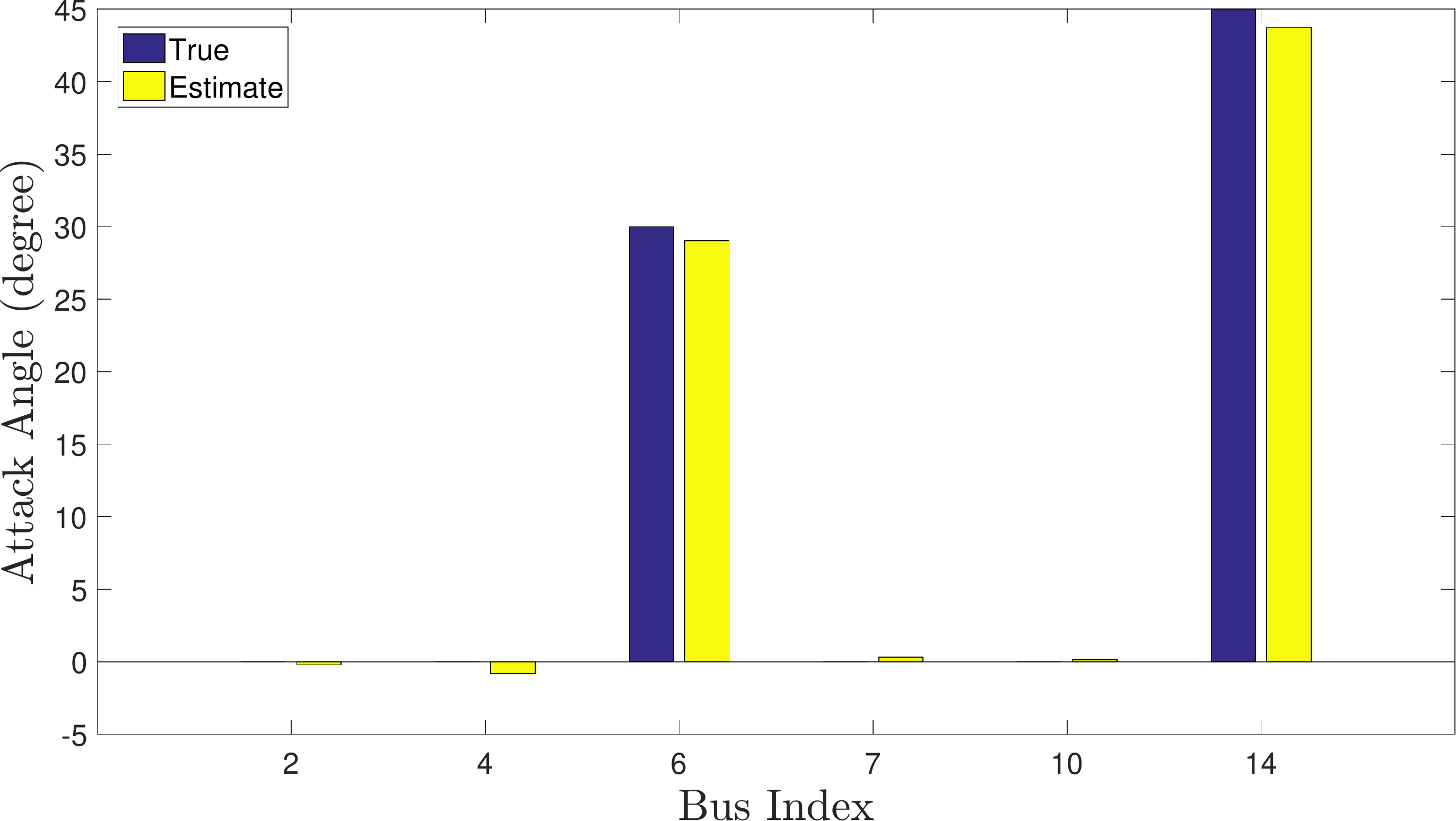}\label{fig:bar_attack_angle}}
\caption{IEEE 14-bus test case: Estimated state (voltage magnitude and phase) and reconstructed attack for attack on two PMUs. (a) True and estimated voltage magnitude. (b) True and estimated voltage phase angle. (c) True and estimated attack angle. }
\label{fig:bar_plots_1}
\end{figure*}

Furthermore, a simultaneous attack on two PMUs is analyzed. The results of Algorithms \ref{alg:Optimal} and \ref{alg:Greedy} are compared. Table~\ref{table:3} indicates that Algorithms \ref{alg:Optimal} and \ref{alg:Greedy} give the same results for the IEEE 14- and 30-bus systems. 

It is important to note the difference in computation time between Algorithms~\ref{alg:Optimal} and~\ref{alg:Greedy}.   
 The difference is more apparent in the 118-bus network than in the 14-bus network, so we report the computation times for the former.\footnote{The algorithms were run on an Intel Xeon E5-1650 v2, 3.5-GHz CPU, 16-GB RAM computer.} Specifically, for simultaneous attack on two PMUs in the 118-bus system,  Algorithm~\ref{alg:Greedy} takes 2 hours and 10 minutes.  On the other hand, Algorithm~\ref{alg:Optimal} required more than 48 hours to solve the total of $\left(\begin{smallmatrix}118 \\ 2\end{smallmatrix}\right)$ nonconvex problems, and the execution was terminated  at 48 hours. 
Thus, Algorithm \ref{alg:Greedy} is computationally efficient.

\begin{table}[!t]
\caption{Vulnerable PMU bus locations for attack on 2 PMUs using Greedy and Optimal algorithms.}
\label{table:3}
 \centering
  \begin{threeparttable}
  \begin{tabular}{c|cc|cc}
    \toprule
    \multirow{2}{*}{Test Case} &
    \multicolumn{2}{c|}{{Algorithm \ref{alg:Optimal} (Optimal)}}  & 
    \multicolumn{2}{c}{{Algorithm \ref{alg:Greedy} (Greedy)}  } \\
      & {$1^{\mathrm{st}}$ PMU} & {$2^{\mathrm{nd}}$ PMU} & {$1^{\mathrm{st}}$ PMU} & {$2^{\mathrm{nd}}$ PMU} \\
      \midrule
    IEEE 14  & 6  & 7 & 6  & 7 \\
    IEEE 30 & 12 & 15 & 12 & 15 \\
    IEEE 118 & {--}\tnote{*} & {--}\tnote{*} & 30 & 40  \\
    \bottomrule
  \end{tabular} 
  \begin{tablenotes}
	\item[*] Computation for 118-bus system was terminated after 48 hours.
  \end{tablenotes}
  \end{threeparttable}
\end{table}

\subsection{SE and Attack Reconstruction with PMU Measurements}
\label{AM_detection}

\begin{figure}[t]
\begin{minipage}[t]{0.48\linewidth}
    \includegraphics[width=\linewidth]{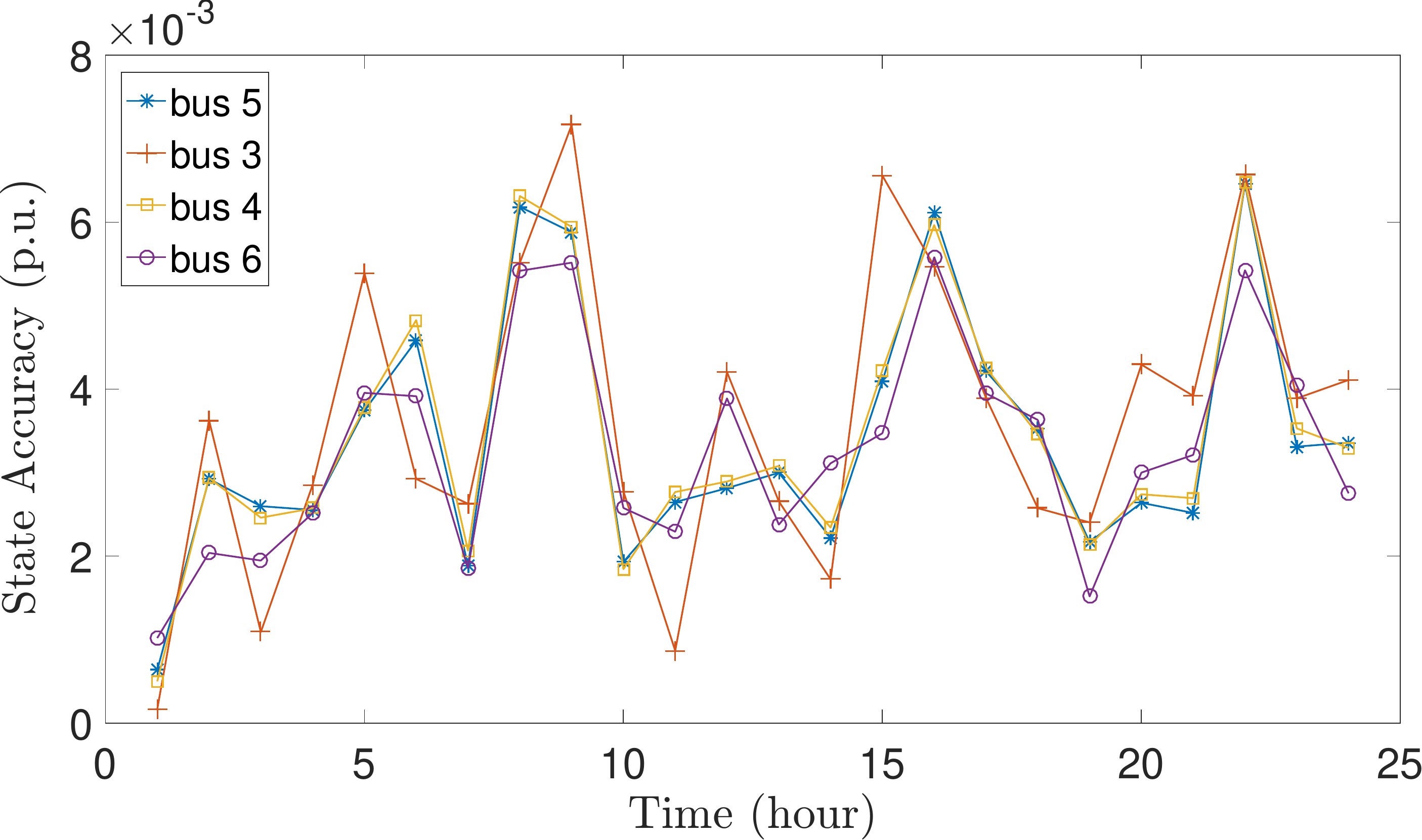}
    \caption{State accuracy for an attack on one PMU as a function of hourly demand in a standard IEEE-118 bus network .}
    \label{fig:plot_demand_state}
\end{minipage}%
    \hfill%
\begin{minipage}[t]{0.48\linewidth}
    \includegraphics[width=\linewidth]{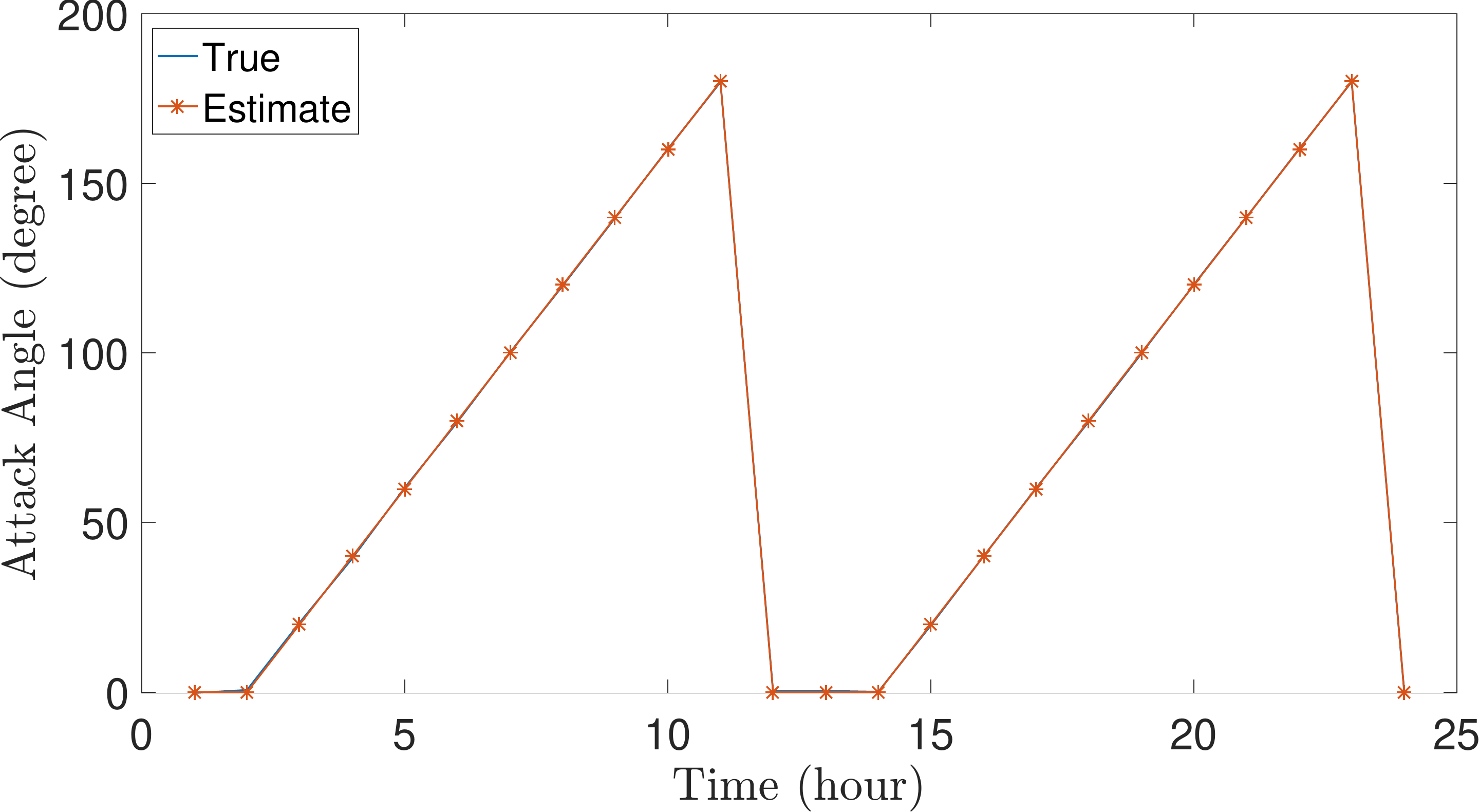}
    \caption{Comparison between the true and estimated attack angle as a function of hourly demand in a standard IEEE-118 bus network.}
    \label{fig:plot_attack_angle_demand}
\end{minipage} 
\end{figure}

\begin{figure}[t]
\begin{minipage}[t]{0.45\linewidth}
    \includegraphics[width=\linewidth]{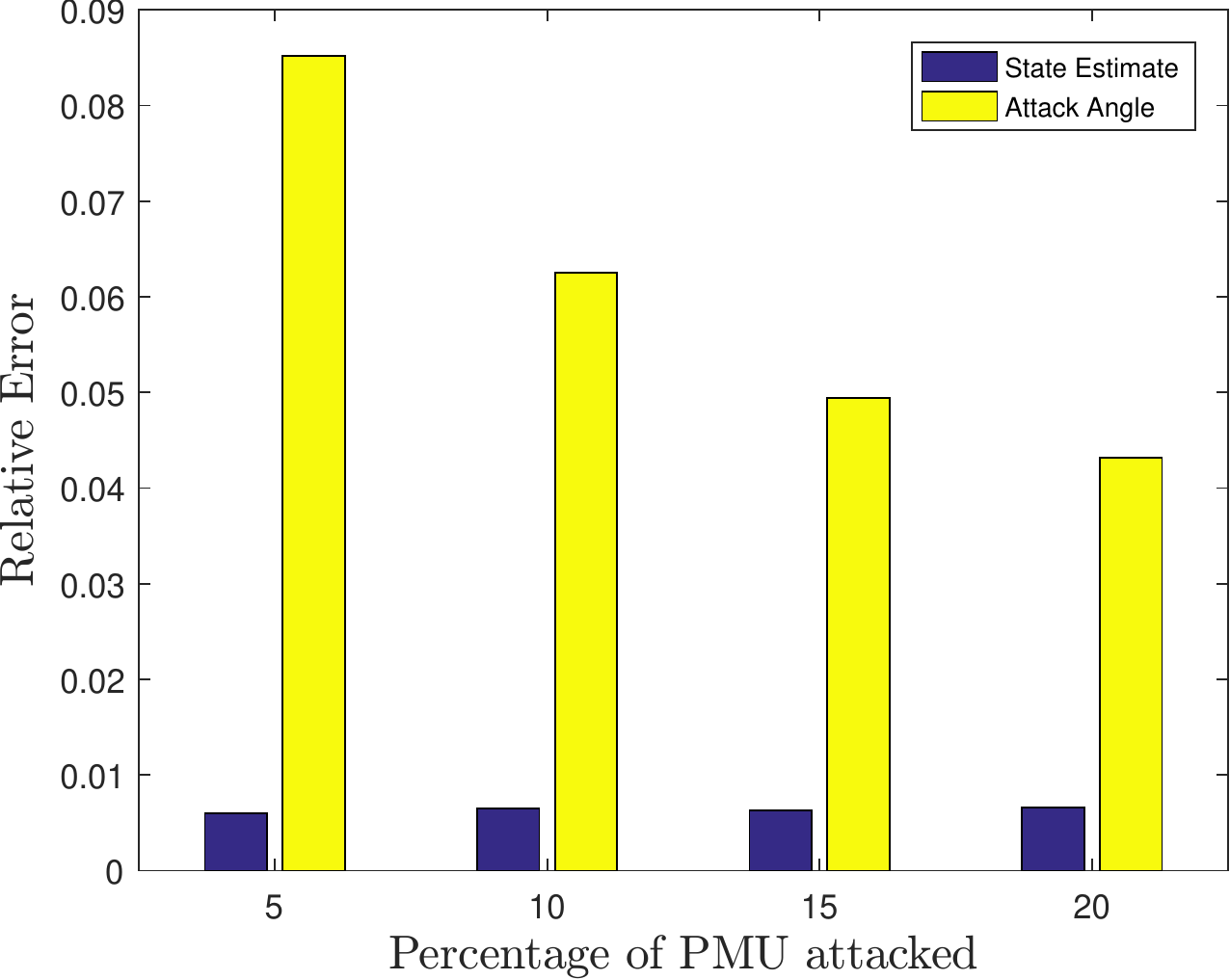}
    \caption{Effect of percentage of PMUs attacked on relative error for a standard IEEE-118 bus network.}
    \label{fig:bar_relative_error_118}
\end{minipage}%
    \hfill%
\begin{minipage}[t]{0.45\linewidth}
    \includegraphics[width=\linewidth]{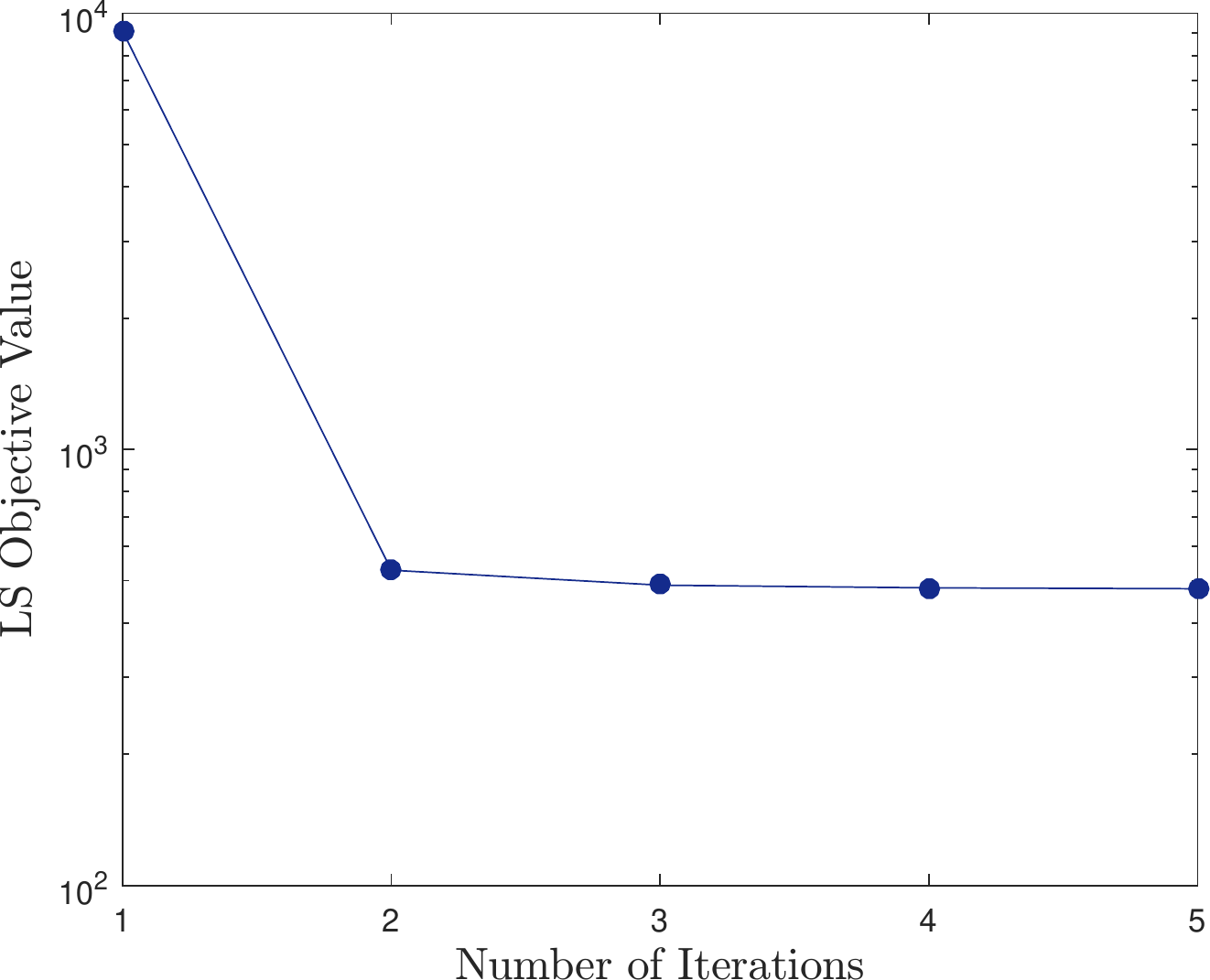}
    \caption{Least Squares (LS) objective value as a function of number of iterations for IEEE-118 system.}
    \label{fig:bar_iterations_118}
\end{minipage} 
\end{figure}

As described in Section \ref{sec:AM}, $\mathbf{z}_n^{\text{atk}}$ is the input to the AM algorithm. The vector $\mathbf{z}_n^{\text{atk}}$ is generated according to~\eqref{eq:znmeas}, where $\mathbf{v}$ is the voltage profile corresponding to nominal network demand obtained from MATPOWER's \texttt{runpf};  $\bm{\Delta\theta}$ is an attack vector which is varied as explained next; and noise vector $\mathbf{w}_n$ is a sample from the Gaussian distribution with zero mean and covariance $\mathbf{\Sigma}_n$ given earlier. 
This renders $\mathbf{z}_n^{\text{atk}}$ random in each run of the algorithm. 
One realization of $\mathbf{z}_n^{\text{atk}}$ with attack on buses 6 and 14 and attack angles $\Delta\theta_6=30^{\circ}$ and $\Delta\theta_{14}=45^{\circ}$ is used to  generate Fig.~\ref{fig:bar_plots_1}. Tolerance $\epsilon=0.01$ was used in the termination criterion  $|\texttt{CurrObj}- \texttt{PrevObj}| / |\texttt{CurrObj}| \leq \epsilon$. The figure depicts the  estimated state (voltage magnitude and phase) as well as reconstructed attack for the IEEE 14-bus network. The figure reveals that the state is correctly estimated, and the attacked buses are identified, together with the attack angle. Notice that the estimated attack angle is almost zero for PMU buses that are not attacked. 

Fig. \ref{fig:plot_demand_state} depicts the state accuracy for an attack on the PMU installed on bus 5 as a function of the hourly demand in the IEEE 118-bus network, where the state accuracy is defined as 
$\sqrt{(\hat{V}_{n,r} - V_{n,r})^2 + (\hat{V}_{n,i} - V_{n,i})^2}$.
In addition to bus 5, the  state accuracy for buses connected to the attacked bus 5 is also plotted. For this test, the November weekday 24-hour demand from \cite{NYSEG} is normalized to 1.5 and is used to scale the nominal demand of the IEEE-118 network, from which $\textbf{v}$ is obtained. 
Attack ranging from $0^{\circ}$ to $180^{\circ}$ is simulated in the time interval of 2 through 11 hours and 14 through 23 hours. Fig. \ref{fig:plot_attack_angle_demand} shows the comparison between the true and estimated attack angle for each hour.

Fig. \ref{fig:bar_relative_error_118} depicts the relative error of the state and attack angle estimates as a function of the percentage of PMUs attacked. The relative state and attack angle estimation errors are defined respectively as
$\frac{||\hat{\mathbf{v}} - \mathbf{v}||_2}{||\mathbf{v}||_2}$ and $\frac{||\widehat{\bm{\Delta\theta}}-\bm{\Delta\theta}||_2}{||\bm{\Delta\theta}||_2}$, where $\hat{\mathbf{v}}$ and $\widehat{\bm{\Delta\theta}}$ represent the state and attack angle vector resulting from Algorithm~\ref{alg:AM Algorithm}, while ${\mathbf{v}}$ and $\bm{\Delta\theta}$ represent their true values. 
For each percentage of attacked PMUs, the relative error averaged over 100 samples of $\mathbf{z}_n^{\text{atk}}$ is depicted. Each realization of $\mathbf{z}_n^{\text{atk}}$ entails a random noise sample, a random set of attacked PMUs for the given percentage, and a random attack angle chosen from a uniform distribution in the interval $[-60^{\circ}, 60^{\circ}]$. The voltage profile $\mathbf{v}$ resulting from the nominal demand in the IEEE-118 network and tolerance of 0.01 are used. Fig.~\ref{fig:bar_relative_error_118} reveals that the relative state estimation error is below 1\% even when 20\% of PMUs are attacked.

\subsection{SE and Attack Reconstruction with SCADA and PMU Measurements}\label{AM_SCADA_detection}

{
As described in Section \ref{sec:PMU_SCADA}, the AM algorithm can be applied to perform the state estimation and attack angle reconstruction with combined PMU and SCADA measurements. MATPOWER's \texttt{run$\_$se.m} is used to obtain $\hat{\mathbf{v}}_s$.
The redundancy ratio for SCADA measurements, which is the ratio of SCADA measurements over the state variables is 2.2 \cite{Monticelli2000}. The SCADA measurements include active and reactive line power flows through ``from'' and ``to'' ends of the bus,  and bus voltage magnitudes. 
For the purpose of this simulation, 50$\%$ of all bus voltage magnitudes and active and reactive line power flows are considered as SCADA measurements, which approximately equals the redundancy ratio. These remain fixed throughout the experiment.  The PMU placement is according to Table~\ref{table:1}, with an additional PMU placed at the slack bus for each network. In order to obtain the relative error, for each network, PMUs on two buses are attacked.
}

The SCADA measurement noise covariance matrix $\bm{\Sigma}_e$ is diagonal resulting from standard deviation of 0.01 and 0.02 for bus voltage magnitude and line power flows (to and from) respectively. The matrix $\bm{\Sigma}_s$ is given by \eqref{eq:SCADA_cov} and obtained from \texttt{run$\_$se.m}.\footnote{MATPOWER's \texttt{run$\_$se.m} is not considering the measurement from the reference bus as unknown for the state estimation; hence, the dimension of $\bm{\Sigma}_s$ returned from  \texttt{run$\_$se.m} is $(2N_b-2) \times (2N_b-2)$. However, the dimension of $\bm{\Sigma}_s$, as described in Section~\ref{sec:PMU_SCADA}, is $(2N_b - 1)\times  (2N_b - 1)$. Thus, we augment $\bm{\Sigma}_s^{-1}$ returned from \texttt{run$\_$se.m}  by one row and one column with a single non-zero entry corresponding to $V_{1,r}$. This entry is set to a value greater in absolute value than the maximum of the remaining entries in $\bm{\Sigma}_s^{-1}$ (approximately $10^8$ here).}

{
The input to the AM algorithm is $\mathbf{z}_n^{\text{atk}}$, which is random due to noise vector $\mathbf{w}_n \sim \mathcal{N}(\mathbf{0}, \bm{\Sigma}_n)$ in the PMU measurement and noise vector $\mathbf{w}_s \sim \mathcal{N}(\mathbf{0}, \bm{\Sigma}_e)$ in the SCADA measurement vector. Thus, the relative error is averaged over 100 realizations of $\mathbf{z}^{\text{atk}}_n$. Tolerance of 0.01 was used for the AM algorithm. Table \ref{table:4} lists the relative state and attack angle estimation errors for the standard test networks in three scenarios. In the first scenario, no SCADA measurements are used. This case is indicated in Table~\ref{table:4} as ``only PMU.'' For the ``only PMU'' scenario, the attacks on the IEEE 14-, 30-, and 118-bus networks are respectively on the PMUs at buses 6 and 14 ($\Delta \theta_6 = 30^{\circ}$, $\Delta \theta_{14} = 45^{\circ}$), 6 and 12 ($\Delta \theta_{6} = 30^{\circ}$, $\Delta \theta_{12} = 45^{\circ}$), and 36 and 50 ($\Delta \theta_{36} = 30^{\circ}$, $\Delta \theta_{50} = 45^{\circ}$).
Along with the SCADA measurements, the second scenario utilizes all the PMUs depicted in Table~\ref{table:1} (in addition to the slack bus). This case is indicated in Table~\ref{table:4} as ``PMU + SCADA.'' For the ``PMU + SCADA'' scenario, the attacks on the IEEE 14-, 30-, and 118-bus networks are respectively on the PMUs at buses 6 and 14 ($\Delta \theta_6 = 30^{\circ}$, $\Delta \theta_{14} = 45^{\circ}$), 12 and 15 ($\Delta \theta_{12} = 30^{\circ}$, $\Delta \theta_{15} = 45^{\circ}$), and 5 and 8 ($\Delta \theta_{5} = 30^{\circ}$, $\Delta \theta_{8} = 45^{\circ}$).
In the third scenario, few PMUs are chosen at random for each network from Table \ref{table:1}; the purpose is to make the corresponding network unobservable (i.e., a singular $\mathbf{G}$) in the absence of SCADA measurements. Hence, this scenario includes the measurements from the reduced set of PMUs and SCADA measurements. This case is indicated in Table \ref{table:4} as ``Reduced PMU + SCADA.'' }

{It is evident from Table~\ref{table:4} that combining PMU and SCADA measurements increases the accuracy of the AM algorithm. Table~\ref{table:4} also indicates that in the Reduced PMU + SCADA scenario, more accurate results are obtained than in the Only PMU scenario.  
}

\begin{table}[!t]
\caption{Relative state and attack angle (A. A.) estimation errors for standard test networks using PMU, PMU + SCADA, and Reduced PMU + SCADA measurements.}
\label{table:4}
 \centering
  \begin{threeparttable}
	  \begin{tabular}{c|cc|cc|cc}
	    \toprule
	    \multirow{3}{*}{Test Case} &
	    \multicolumn{2}{c|}{Only PMU}  & 
	    \multicolumn{2}{c|}{PMU + SCADA} &
	    \multicolumn{2}{c}{Reduced + SCADA} \\
		&   &  &  &  & PMU  \\
		& {State} & {A. A.} & {State} & {A. A.} & {State} & {A. A.} \\
	      \midrule
	    IEEE 14  & 0.0210  & 0.0577 & 0.0055 & 0.0258 & 0.0066 & 0.0202 \\
	    IEEE 30 & 0.0970 & 0.3727 & 0.0091 & 0.0519 & 0.0103 & 0.0379 \\
	    IEEE 118 & 0.0073 & 0.1213 & 0.0010 & 0.0938 & 0.0013 & 0.0689  \\
	    \bottomrule
	  \end{tabular} 
  \end{threeparttable}
\end{table}

\subsection{Comparision between Algorithm \ref{alg:AM Algorithm} and the LNRT}\label{Bad_data}

In power systems, bad data can arise from multiple sources such as corrupted meter measurements, communication failures, and parameter uncertainty.  This section examines whether a classical  bad data detector can identify the GPS-spoofed PMU measurements and return a highly accurate state estimate. The LNRT is selected to this end, which is capable to detect and identify bad data from the measurement residuals, as opposed to the chi-square test, which can detect the presence of bad data, but not identify their locations. 

To streamline the notation, the measurement equation for LNRT is $\mathbf{z} = \mathbf{H} \mathbf{v} + \mathbf{w}$, where $\mathbf{z}$ and $\mathbf{H}$ are formed by stacking $\{\mathbf{z}_n\}_{n \in \mathcal{N}_{\text{PMU}}}$ and $\{\mathbf{H}_n\}_{n \in \mathcal{N}_{\text{PMU}}}$ respectively. Furthermore, the noise $\mathbf{w}$ has a block diagonal covariance matrix $\bm{\Sigma}$ with diagonal blocks $\{\bm{\Sigma}_n\}_{n \in \mathcal{N}_{\text{PMU}}}$. The residual of the measurement vector is given by $\mathbf{r} = \mathbf{z} - \mathbf{H} \hat{\mathbf{v}}$ where $\hat{\mathbf{v}} = {(\mathbf{H}^{\top}\bm{\Sigma}^{-1}\mathbf{H})}^{-1}\mathbf{H}^{\top}\bm{\Sigma}^{-1}\mathbf{z}$ is computed via weighted least squares. The LNRT uses the residual statistics to detect and identify bad data. In particular, the residual vector has distribution $\mathbf{r} \sim \mathcal{N}(\mathbf{0}, \bm{\Omega})$, where $\bm{\Omega} = \mathbf{S} \bm{\Sigma}$ and $\mathbf{S} = \mathbf{I} - \mathbf{H} (\mathbf{H}^{\top} \bm{\Sigma}^{-1} \mathbf{H})^{-1} \mathbf{H} \bm{\Sigma}^{-1}$. 

The LNRT uses the weighted least squares to obtain the state estimate ($\hat{\mathbf{v}}$) and measurement residual ($\mathbf{r}$). It utilizes the normalized measurement residuals to identify bad data. The normalized residual $\mathbf{r}_\text{normalized} = \mathbf{r} /\sqrt{{\Omega}_{i,i}}$ should follow the standard normal distribution for all $i$ when bad data are absent. If $\mathbf{r}_\text{normalized}$ is larger than a threshold, chosen here as 3~\cite[Sec.~4.8.4]{Antonio2008}, then the particular measurement is removed.\footnote{Critical measurements are not removed, because elimination of a critical measurement would render the system unobservable.} Then the least squares solution is re-computed, and the process continues until all bad data have been removed.

In this experiment, PMUs are placed as listed in Table~\ref{table:1}, and the AM algorithm as well as the LNRT are run on 200 realizations of $\mathbf{z}_n^{\text{atk}}$. Tolerance of 0.0001 was used for Algorithm~\ref{alg:AM Algorithm}. For the IEEE 14-, 30-, and 118-bus networks, the attacks are respectively on the PMUs at buses 2 and 14 ($\Delta \theta_2 = 60^{\circ}$, $\Delta \theta_{14} = 70^{\circ}$), 11 and 12 ($\Delta \theta_{11} = 70^{\circ}$, $\Delta \theta_{12} = 60^{\circ}$), and 64 and 2 ($\Delta \theta_{64} = 70^{\circ}$, $\Delta \theta_{2} = 70^{\circ}$). 
Note that the GPS-spoofing attack not only affects the voltage measurement at the attacked bus but also the current measurements flowing on the lines connected to the attacked bus. Our observation is that the LNRT can identify the voltage and some of the current measurements corresponding to the attacked bus as bad, but in many cases, it cannot identify all affected current measurements.

Table \ref{table:5-1} compares the relative state error (averaged over the 200 realizations) returned by Algorithm~\ref{alg:AM Algorithm} and by LNRT. Although LNRT removes multiple bad measurements, it can be observed from Table \ref{table:5-1} that Algorithm \ref{alg:AM Algorithm} produces more accurate results than these obtained from LNRT. These tests indicate the strength of the novel AM algorithm. 
Table \ref{table:5-2} compares the computation time of Algorithm~\ref{alg:AM Algorithm} and LNRT for the standard test networks.\footnote{The system specifications for the computation times of Tables~\ref{table:5-2} and~\ref{table:6-2} are as follows: Intel Xeon E3-1271 v3, 3.6-GHz CPU, 32-GB RAM. The computation times include averaging over the 200 realizations.}
It is finally worth noting that the AM algorithm can also yield relatively accurate estimates of the attack angles, contrary to LNRT.

\begin{table}[t]
\caption{Relative state estimation errors for standard test networks using Algorithm \ref{alg:AM Algorithm} and the LNRT.}
\label{table:5-1}
 \centering
  \begin{threeparttable}
  \begin{tabular}{c|cc|cc}
    \toprule
    \multirow{2}{*}{Test Case} &
    \multicolumn{2}{c|}{Attack on One PMU}  &
    \multicolumn{2}{c}{Attack on two PMUs} \\
      & {Algorithm \ref{alg:AM Algorithm}} & {LNRT} & {Algorithm \ref{alg:AM Algorithm}} & {LNRT}  \\
      \midrule
    IEEE 14  & 0.0142 & 0.0531 & 0.0145 & 0.1536 \\
    IEEE 30 & 0.0469 & 0.0539 & 0.0566 & 0.0835 \\
    IEEE 118 & 0.0039 & 0.0054 & 0.0039 & 0.0057  \\
    \bottomrule
  \end{tabular} 
  \end{threeparttable}
\end{table}

\begin{table}[t]
\caption{Computation time (in seconds) for Algorithm \ref{alg:AM Algorithm} and the LNRT in standard test networks.}
\label{table:5-2}
 \centering
  \begin{threeparttable}
  \begin{tabular}{c|cc|cc}
    \toprule
    \multirow{2}{*}{Test Case} &
    \multicolumn{2}{c|}{Attack on One PMU}  &
    \multicolumn{2}{c}{Attack on two PMUs} \\
      & {Algorithm \ref{alg:AM Algorithm}} & {LNRT} & {Algorithm \ref{alg:AM Algorithm}} & {LNRT}  \\
      \midrule
    IEEE 14  & 6.8063 & 0.7324 & 8.4220 & 0.9419 \\
    IEEE 30 & 49.5106 & 0.9430 & 91.6521 & 1.9756 \\
    IEEE 118 & 95.8400 & 106.4569 & 102.8687 & 150.6190  \\
    \bottomrule
  \end{tabular} 
  \end{threeparttable}
\end{table}

\subsection{Comparision between Algorithm \ref{alg:AM Algorithm} and the Spoofing-Matched Algorithm (SpM)}\label{SpM}

The spoofing-matched algorithm (SpM) is developed in~\cite{Fan2017}, where it is demonstrated to successfully detect and correct a single GPS spoofing attack. In this section, we compare the performance of Algorithm~\ref{alg:AM Algorithm} with SpM under attacks on two PMUs. 

In this experiment, two PMUs from each network listed in Table~\ref{table:1} are attacked. 
For the IEEE 14-, 30-, and 118-bus networks, the attacks are respectively on the PMUs at buses 6 and 7 ($\Delta \theta_6 = \Delta \theta_{7} = 90^{\circ}$), 6 and 10 ($\Delta \theta_{6} = \Delta \theta_{10} = 90^{\circ}$), and 3 and 4 ($\Delta \theta_{3} = \Delta \theta_{4} = 90^{\circ}$).
 The AM algorithm as well as the SpM are run on 200 realizations of $\mathbf{z}_n^{\text{atk}}$.
The tolerance is set to 0.0001 for Algorithm \ref{alg:AM Algorithm}. 
 The SpM algorithm utilizes the golden section search algorithm~\cite[Chapter 7]{Optimization_book} to determine the estimate of the GPS spoofed attack angle. The accuracy of the golden section algorithm depends on the difference between the lower and upper bound of an interval. In other words, if the the difference is less than the desired tolerance, convergence is declared. Here, the tolerance of the golden section algorithm is set to $10^{-5}$.

Table \ref{table:6-1} summarizes the performance of both algorithms.
The table shows that in terms of relative state and attack angle estimation errors, Algorithm \ref{alg:AM Algorithm} performs better than SpM. 
Finally, Table~\ref{table:6-2} compares the computation time between Algorithm~\ref{alg:AM Algorithm} and SpM for the standard test networks. Note that the computation times of Algorithm~\ref{alg:AM Algorithm} reported in Tables~\ref{table:5-2} and~\ref{table:6-2} are slightly different because the attack locations and angles are different between the two experiments. 

\begin{table}[t]
\caption{Relative state and attack angle (A. A.) estimation errors for standard test networks using Algorithm \ref{alg:AM Algorithm} and the SpM.}
\label{table:6-1}
 \centering
 \begin{threeparttable}
  \begin{tabular}{c|cc|cc}
    \toprule
    \multirow{2}{*}{Test Case} &
    \multicolumn{2}{c|}{Algorithm \ref{alg:AM Algorithm}}  &
    \multicolumn{2}{c}{SpM} \\
      & {State} & {A.A.} & {State} & {A.A.}  \\
      \midrule
    IEEE 14  & 0.0143 & 0.0172 & 0.2875 & 0.7187 \\
    IEEE 30 & 0.0550 & 0.0904 & 0.1226 & 0.7087 \\
    IEEE 118 & 0.0038 & 0.0427 & 0.0170 & 0.7072  \\
    \bottomrule
  \end{tabular} 
  \end{threeparttable}
\end{table}

\begin{table}[!t]
\caption{Computation time (in seconds) for Algorithm \ref{alg:AM Algorithm} and the SpM algorithm in standard test networks.}
\label{table:6-2}
\centering
\begin{tabu} to 0.5\textwidth {  X[c] | X[c] | X[c] }
\hline
 Test Case  & Algorithm \ref{alg:AM Algorithm} & SpM  \\
 \hline
 IEEE 14 & 9.0527 & 8.4909\\

 IEEE 30 & 110.5931 & 45.0878 \\

 IEEE 118  & 91.3241 & 1.2191$\cdot 10^4$ \\
\hline
\end{tabu}
\end{table}

\section{Summary and Future Directions}\label{sec:Conclusion}
In this paper, we build on the novel measurement model developed in \cite{Risbud2016} to formulate an optimization problem to identify the most vulnerable PMUs in the network. A greedy algorithm is developed to approximately solve the aforementioned problem. The problem of jointly estimating the network state and reconstructing the attack is cast as a nonconvex constrained least squares problem, and an alternating minimization algorithm is developed for its solution. Thorough numerical tests are performed to compare the performance of the developed algorithms (in terms of accuracy, detection, and computational time) with other state-of-the-art methods. Numerical tests illustrate the potential of the developed algorithms in this paper.

Building on the developed approach in this paper, it is of interest to perform simultaneous extraction of the attack and dynamic state estimation. We plan to utilize dynamic observers for nonlinear systems with time-delays to this end.

\bibliographystyle{IEEEtran}

\bibliography{TSG_citation}

\begin{IEEEbiography}
[{\includegraphics[width=1in,height=1.25in,clip,keepaspectratio]{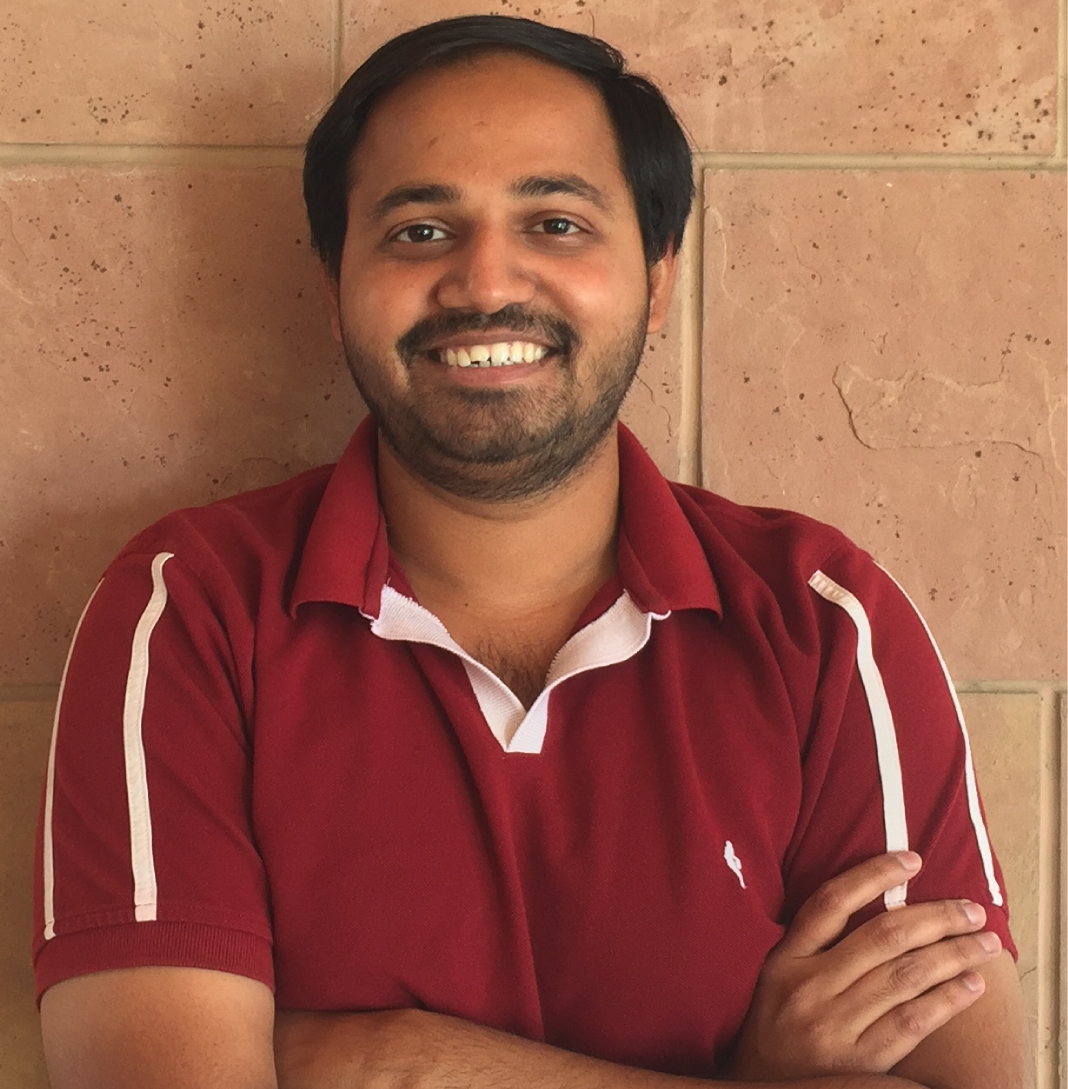}}]{Paresh Risbud} 
received the B.E in Electronics Engineering from the University of Mumbai, India in 2011. He completed the M.S. degree in Electrical and Computer Engineering from the University of Texas at San Antonio in 2014. Currently, he is working towards the Ph.D. degree in the department of Electrical and Computer Engineering at the University of Texas at San Antonio, where he is a graduate research assistant. 

His research interests are in the area of statistical signal processing, power system state estimation, and optimization and control of cyber-physical systems.
\end{IEEEbiography} 

\begin{IEEEbiography}
[{\includegraphics[width=1in,height=1.25in,clip,keepaspectratio]{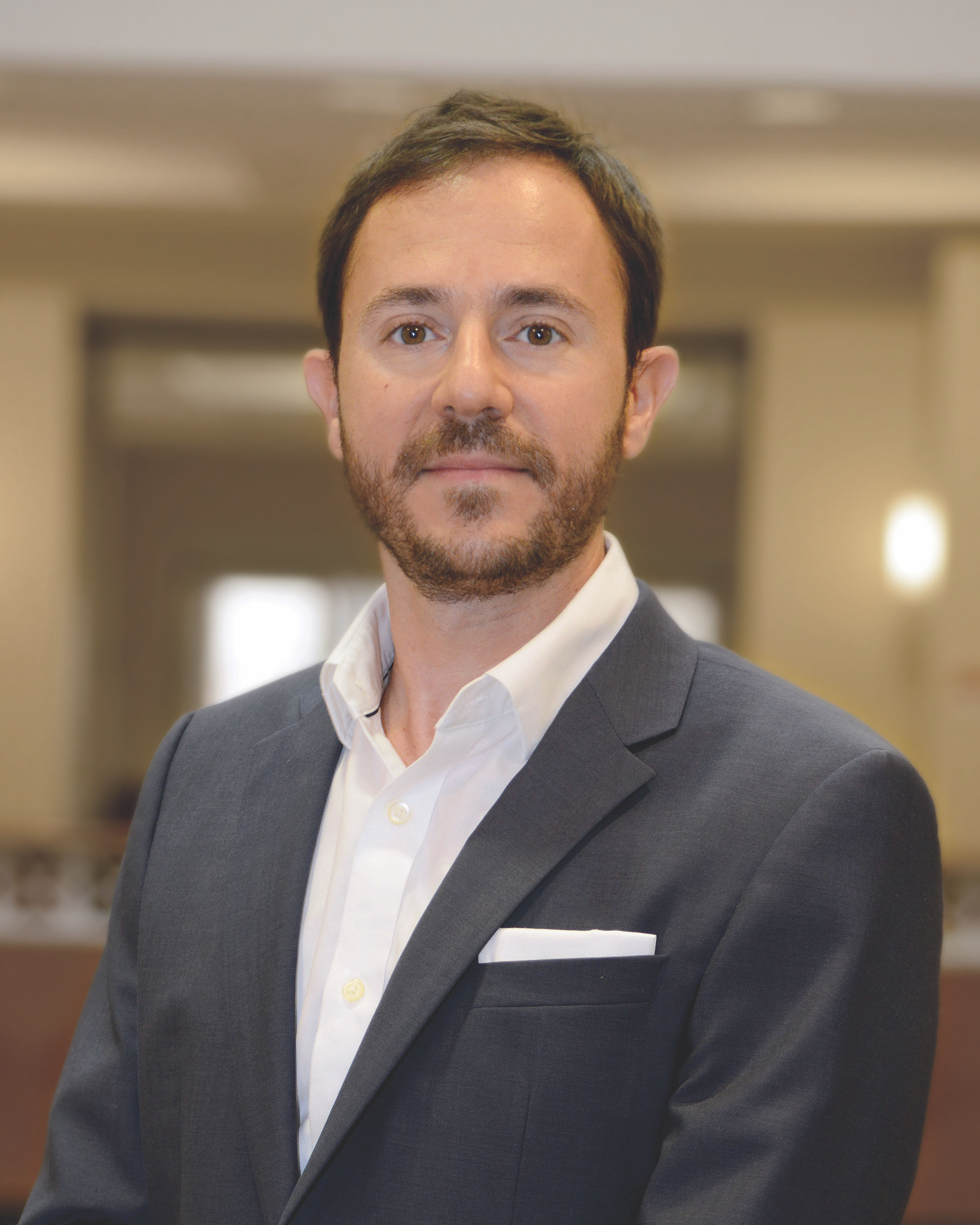}}]
{Nikolaos Gatsis}
received the Diploma degree in Electrical and Computer Engineering from the University of Patras, Greece, in 2005 with honors. He completed his graduate studies at the University of Minnesota, where he received the M.Sc. degree in Electrical Engineering in 2010, and the Ph.D. degree in Electrical Engineering with minor in Mathematics in 2012. 

He is currently an Assistant Professor with the Department of Electrical and Computer Engineering at the University of Texas at San Antonio. His research focuses on optimal and secure operation of smart power grids and other critical infrastructures, including water distribution networks and the Global Positioning System. 

Dr. Gatsis has co-organized symposia in the area of
smart grids in IEEE GlobalSIP 2015 and 2016. He has
also served as a co-guest editor for a special issue of the IEEE Journal
on Selected Topics in Signal Processing on critical infrastructures.
\end{IEEEbiography} 

\begin{IEEEbiography}
[{\includegraphics[width=1in,height=1.25in,clip,keepaspectratio]{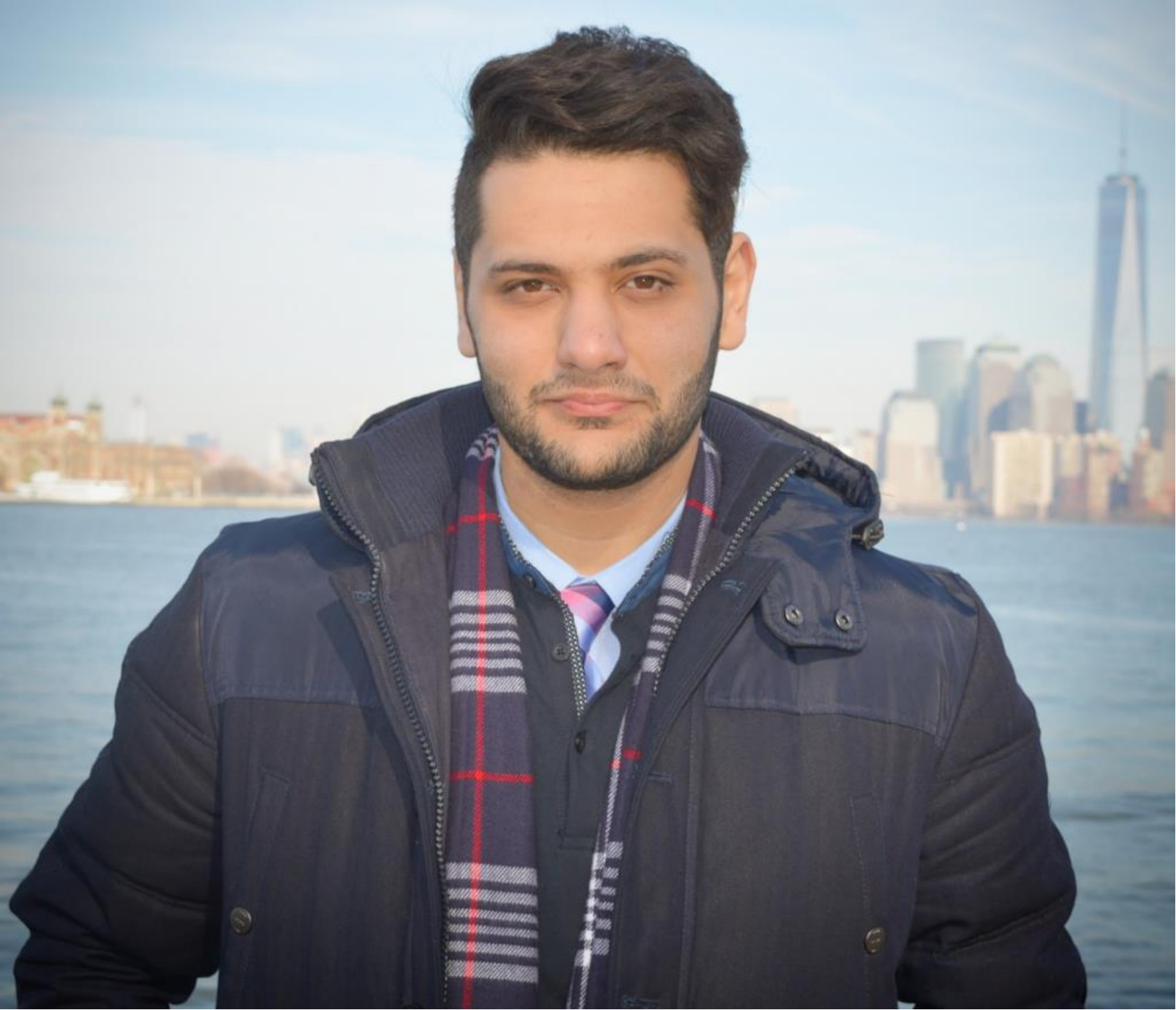}}]
{Ahmad Taha}
(S'07--M'15) received the B.E. and Ph.D. degrees in Electrical and Computer Engineering from the American University of Beirut, Lebanon in 2011 and Purdue University, West Lafayette, Indiana in 2015. 

In Summer 2010, Summer 2014, and Spring 2015 he was a visiting scholar at MIT, University of Toronto, and Argonne National Laboratory. Currently he is an assistant professor with the Department of Electrical and Computer Engineering at The University of Texas, San Antonio. Dr. Taha is interested in understanding how complex cyber-physical systems operate, behave, and \textit{misbehave}. His research focus includes optimization and control of power system, observer design and dynamic state estimation, and cyber-security.
\end{IEEEbiography}

\end{document}